# Accounting for Randomness in Measurement and Sampling in Study of Cancer Cell Population Dynamics

Siavash Ghavami[1*,**], Olaf Wolkenhauer[2**,***], Farshad Lahouti[3*], Mukhtar Ullah[4**], Michael Linnebacher[5****]

**Abstract.** Studying the development of malignant tumours, it is important to know and predict the proportions of different cell types in tissue samples. Knowing the expected temporal evolution of the proportion of normal tissue cells, compared to stem-like and non-stem like cancer cells, gives an indication about the progression of the disease and indicates the expected response to interventions with drugs. Such processes have been modeled using Markov processes. An essential step for the simulation of such models is then the determination of state transition probabilities. We here consider the experimentally more realistic scenario in which the measurement of cell population sizes is noisy, leading to a particular hidden Markov model. In this context, randomness in measurement is related to noisy measurements, which are used for the estimation of the transition probability matrix. Randomness in sampling, on the other hand, is here related to the error in estimating the state probability from small cell populations. Using aggregated data of fluorescence-activated cell sorting (FACS) measurement, we develop a minimum mean square error estimator (MMSE) and maximum likelihood (ML) estimator and formulate two problems to find the minimum number of required samples and measurements to guarantee the accuracy of predicted population sizes using a transition probability matrix estimated from noisy data. We analyze the properties of two estimators for different noise distributions and prove an optimal solution for Gaussian distributions with the MMSE. Our numerical results show, that for noisy measurements the convergence mechanism of transition probabilities and steady states differ widely from the real values if one uses the standard deterministic approach in which measurements are assumed to be noise free. This provides support for our argument that for the analysis of FACS data one should consider the observed state a random variable. The second problem we address is about the consequences of estimating the probability of a cell being in a particular state from measurements of a population of cells. For small population sizes the law of large numbers will not be satisfied. We show how the uncertainty arising from small sample sizes can be captured by a distribution for the state probability rather than a single value. Our work contributes to a better understanding of randomness and noise when studying stochastic phenomena using FACS data.

Emails: [1]s.ghavami@ut.ac.ir [2]olaf.wolkenhauer@uni-rostock.de, [3]lahouti@ut.ac.ir, [4]mukhtar.ullah@nu.edu.pk, [5]michael.linnebacher@med.uni-rostock.de
\* Center for Wireless Multimedia Communications, Center of Excellence in Applied Electromagnetic Systems, School of Electrical & Computer Engineering, College of Engineering, University of Tehran, Tehran, Iran, wmc.ut.ac.ir
\*\* Department of Systems Biology and Bioinformatics, University of Rostock, Rostock, Germany, www.sbi.uni-rostock.de.
\*\*\* Stellenbosch Institute for Advanced Study (STIAS), Wallenberg Research Centre at Stellenbosch University, Stellenbosch, South Africa.
\*\*\*\* Department of General, Thoracic, Vascular and Transplantation Surgery, University of Rostock, Rostock, Germany, http://www.moi.med.uni-rostock.de/

# 1. Introduction

The development of cancer or cancerous tumors is studied by observing the proportions of different cell types in tissue samples, changing over time as a consequence of cell proliferation. The notion of cancer stem cells (CSCs) exists for over a century [1-5], but only appeared at the front of cancer research with the identification of molecular markers that allowed the isolation of leukemic CSCs [1, 6]. Recent reviews have discussed important topics in CSC biology, such as the tumor cell of origin [7], therapy resistance [8] and the role of the immune system [1, 9]. A central characteristic of CSCs is their potential for self-renewal and multi lineage differentiation [1, 2, 8]. For example, using the CSC model, minimal residual disease and tumor recurrence after treatment would result from a remaining, therapy-resistant CSC fraction, whereas metastatic potential would be a CSC-specific property [1, 3, 5]. The hypothesis of CSCs leading to tumors is conceptually attractive but requires further research to be confirmed. The CSCs form a (very) small proportion of the tumor and the theory suggests that conventional chemotherapies kill differentiated or differentiating cells, which however form the bulk of the tumor, but are unable to generate new cells. In order to develop optimal therapies it is thus essential to know and predict the temporal evolution of proportions of different cell types in tissue samples.

Modeling proliferation in populations of cells is challenging because the intrinsic biological randomness [10] is combined with extrinsic randomness arising from experimental measurements using markers to detect cell states [11]. While many processes in biology exhibit the Markov property remarkably well [12], if the number of measurements (samples) is limited and the measurements of the counts of cells are noisy, the estimation of transition probabilities for a Markov process is challenging. For populations of cells, the Markov property implies that an initial inhomogeneous population of cells converges to a steady state or oscillatory configuration. We show that conventional Markov models and deterministic approaches to estimate state transition probabilities fail to provide accurate predictions if state measurements are noisy. We here refer to the 'deterministic approach' as the scenario without measurement or sampling noise, which allows for optimal solution to the estimation of elements of the transition probability matrix through solving a system of linear equations. As demonstrated in the present paper, a solution is to consider a hidden Markov model in which two types of randomness, due to measurement and sampling, are taken into account.

In [13-17] the influence of intrinsic and extrinsic noise on processes occurring at the intracellular and cell population levels is investigated. A linear-noise approximation was used to obtain intrinsic noise statistics of biochemical networks in [13, 14]. In [15, 16] the statistical properties of molecular movement is analyzed in crowded intracellular compartments, where the crowding originates from various inert macromolecules in the path of mobile reactant molecules. In [17], it was found that cell-cell variability stemming from both intrinsic and extrinsic sources of noise influences pattern formation at the cell population level, thus showing the importance of noise in developmental biology. Furthermore, in [18, 19], the effects of background light and electronic noises on the overall signals the FACS instrument obtains for each cell is considered.

In [20, 21], the transition probability matrix of a Markov chain is estimated from aggregated data. When the data is noise free, estimating the transition probability matrix of a Markov process with $M$ states requires only $M^2$ measurements. In fact, by solving a system of linear equations, the transition probability matrix of the underlying Markov process can be calculated. Such a deterministic approach was used in [10] for calculating the transition probability matrix of a Markov process describing the proliferation of breast cancer cells. What we would like to point out in the present paper is that when the aggregated data is noisy, the calculation of transition probabilities using the deterministic approach dramatically degrades. Moreover, here using standard algorithms for estimation of the transition probability matrix in a hidden Markov model such as the Baum-Welch algorithm [22] is not possible due to the fact that as we shall see, only noisy versions of state probabilities can be observed.

Fig. 1 shows the normalized histogram of stem like cell populations, which is obtained from experimental data of two different samples. The experiment setup is presented in Section A of the Supplementary Materials. The number of stem like cells in the steady state regime is a random variable. One may argue that this approximately follows a Gaussian distribution. For our study, we will also consider other noise types to compare estimators.

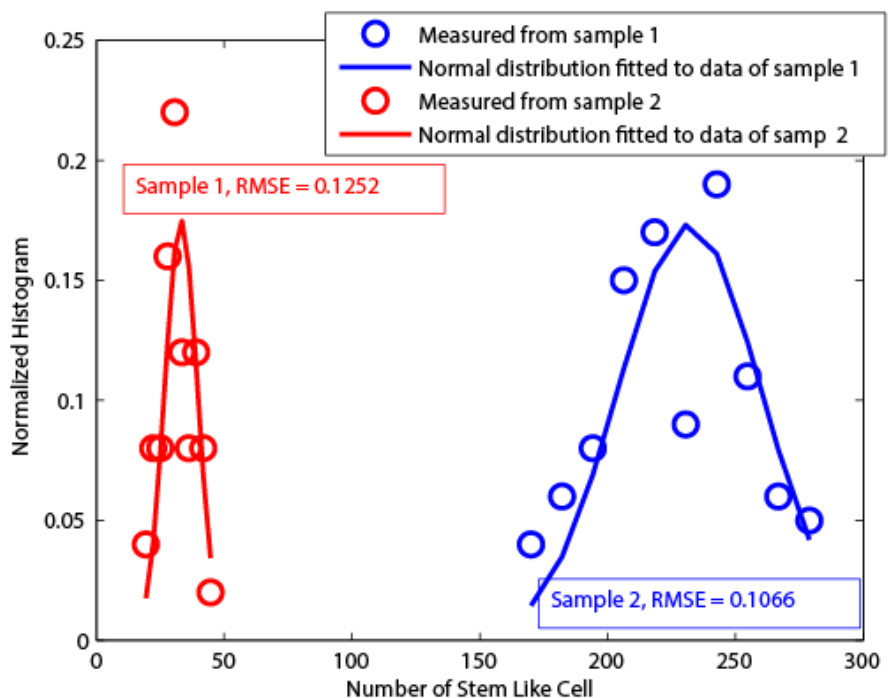

Fig. 1. Normalized histogram plot of cell types in experimental tissue samples of two different patients, showing that to a rough degree of approximation the distributions of data are Gaussians. RMSE stands for Root Mean Square Error and is used as a fitting criterion. The data, that motivated the present work where obtained from HROC87 (red) and HORC24 (blue), two cell lines recently established from the primary colorectal cancers of two patients [23]. Of note, the analyses were performed in very low passages and thus in very high proximity to the behavior of the primary tumor cells. The data for these normalized histograms are reported in Section A of Supplementary Materials.

Here, we consider the effect of measurement and sampling randomness on the prediction of the cell population sizes using an estimated transition probability matrix. The transition probability denoted by $p_{ij}$, is defined between the cell type (state) $i, i \in \{1,...,M\}$, and the cell type (state) $j, j \in \{1,...,M\}$. The observation we wish to make from each cell type is their population size or alternatively their proportion among all cell types (state probability) in a given time instant. One of randomness sources is noisy measurements of cell population sizes by fluorescence activated cell sorting (FACS) techniques. Imperfect cell counting in FACS adds error in estimating the number of cells during proliferation. Statistically, one of the measurement noise sources in cell counting is shot noise [24], which typically amounts to a Poisson distribution [25]. Moreover, our experimental measurements show that for FACS analyses the noise distribution is fitted to Gaussian distribution if the number of cells is reasonably large. The number of cells in a population is always positive, which is why the Gaussian assumption would have to be replaced by the Poisson distribution for small numbers of cells. The randomness in sampling is due to estimation errors for the probability of states, linked to the law of large numbers (LLN) used for estimation of state probabilities of processes with small population sizes.

For improved accuracy of an estimated transition probability matrix, we use minimum mean square error (MMSE) and maximum likelihood (ML) estimators. Our simulation results show that the error in counting the population size of each cell type dramatically degrades the accuracy of the estimated transition probability matrix when deterministic estimation approaches are used. The principal goal of this work is to determine the minimum number of measurements that is required in order to guarantee the accuracy of population sizes predicted using a transition probability matrix, estimated based on noisy data. The main challenge for this problem is the small number of samples that is typically available from FACS analyses in systems biology. We consider the error in prediction of the size of each cell type population and the average error in prediction of all population sizes as two possible measures of accuracy, which we aim to limit in two design problems we formulate. Table 1 summarizes different types of uncertainty in FACS data analysis for the prediction of cell population sizes including 1) imperfect markers [11], 2) imperfect sorting [11], 3) imperfect counting of cell population sizes and 4) imperfect state probability estimation. The types and sources of uncertainty are described and it is also noted how each uncertainty type is dealt with in the sequel.

The remainder of this paper is organized as follows. In Section 2 the system model and problem statements are presented. Section 3 reports the stochastic analyses. Simulation results are presented in Section 4. Finally conclusions are drawn in Section 5.

Table 1. Types and sources of uncertainty of FACS data and data analyses in prediction of cell population sizes.

| Type of uncertainty | Source of uncertainty | Description | Reference |
|---|---|---|---|
| Imperfect marker | Markers used to categorize cell types are frequently imperfect [1]. | In the imperfect marker model, it is assumed that the marker does not allow to sort all the cancer stem cells [1] but can at most separate the cells into CSC-rich and CSC-poor populations. | [11] |
| Imperfect sorting | Some cells are assigned to the wrong category due to errors of FACS | Sorting by FACS typically involves errors: some cells are assigned to the wrong category. | [11] |
| Imperfect counting of cell population sizes | Noise of measurement | The number of cells in each cell state is erroneous due to the noise in counting. The uncertainty in measurement subsequently leads to an erroneous transition probability matrix. | An MMSE or ML estimator is derived for estimation of the transition probability matrix. This estimator considers the statistics of noise. |
| Imperfect state probability estimation | Using LLN for estimation of cell probability | Motivated by LLN, the state probability for each cell is estimated by measuring the number of cells in each state and by dividing it by the total population size. When the population size is not large enough LLN cannot be used for estimating cell state probability and it should be considered as a random variable. | Description of cell state probability is made using a probability density function instead of a single value. |

## 2. System Model & Problem Statement

### 2.1. System Model

To design novel therapies for cancer treatment, detailed knowledge about the dynamics of cancer cell growth is necessary. It is demonstrated in [1, 10] that cell growth dynamics can be explained by a first order Markov model in which cells transit stochastically between different states. A second prediction in [10] is that breast cancer stem-like cells arise *de novo* from non-stem-like cells. These findings contribute to our understanding of cancer heterogeneity and reveal how stochasticity in single-cell behaviors promotes phenotypic equilibrium in populations of cancer cells. Hence, estimating the transition probability matrix of this Markov process has a fundamental role in describing the dynamic process of stem-like and non-stem-like cell growth. The cell population counting is done using FACS. Imperfect counting in FACS imposes randomness on the measured number of initial cells [26-30]. We will refer to this randomness as measurement noise in the remainder of this paper. Fig. 2 shows the effect of sorting and counting on cell population sizes of parental SUM159 of breast cancer consisting of stem-like, basal and luminal cells [10]. The basal cells and luminal cells form respectively, the outer and the inner layers of the glandular tissue of the breast [31]. Specific molecular markers are used to identify different cell types. Cells are sorted according to different populations and the cell populations are counted. Ideally, this should be performed after each doubling time with perfect (without noise of counting) and imperfect counting (with noise of counting). The cell population size using imperfect counting is used for state probability estimation and consequently for transition probability estimation, which can be modeled with hidden Markov model. Fig. 2 schematically demonstrates the effect of imperfect counting after one cell doubling. This is one source of error in estimation of the transition probability matrix.

In this manuscript it is assumed that the probability density function (PDF) of measurement noise is known, which is obtained from our experimental measurements (reported in Fig. 1) and PDF of shot noise. If the number of cells is small, shot noise in particular can add significant uncertainty to the exact values of cell numbers [32, 33]. The statistical characteristics of this noise are described by a Poisson distribution [25]. Moreover, normalized histogram plot of cell types in experimental tissue samples of two different patients is reported in Fig. 1. It is shown that the noise in cell counting, with a rough degree of approximation, may be described by a Gaussian distribution. When performing such FACS-analyses, one thus tries to measure a random process (with unknown statistical moments) with a noisy measurement approach. Moreover, it is assumed that, the initial number of different cancer cell types differ in different tissues or cell samples at the start of each experiment. Also, no information is available about the statistics of the initial population of each cell type in the analysis. The transition probability matrix between different cell types is then given by

$$\mathbf{P} = \begin{bmatrix} p_{11} & \cdots & p_{1M} \\ \vdots & \ddots & \vdots \\ p_{M1} & \cdots & p_{MM} \end{bmatrix}, \tag{1}$$

where $M$ is the number of states, which corresponds to the number of recognizable cell types in the experiment. Increasing the number of samples analyzed in the experiment can reduce the effect of noisy measurements but it cannot directly control the randomness of the initial cell population. Hence, as a first step, we try to minimize the cost of experiments by determining the number of samples and measurements to be analyzed, given that a specified level of accuracy of cell population is needed.

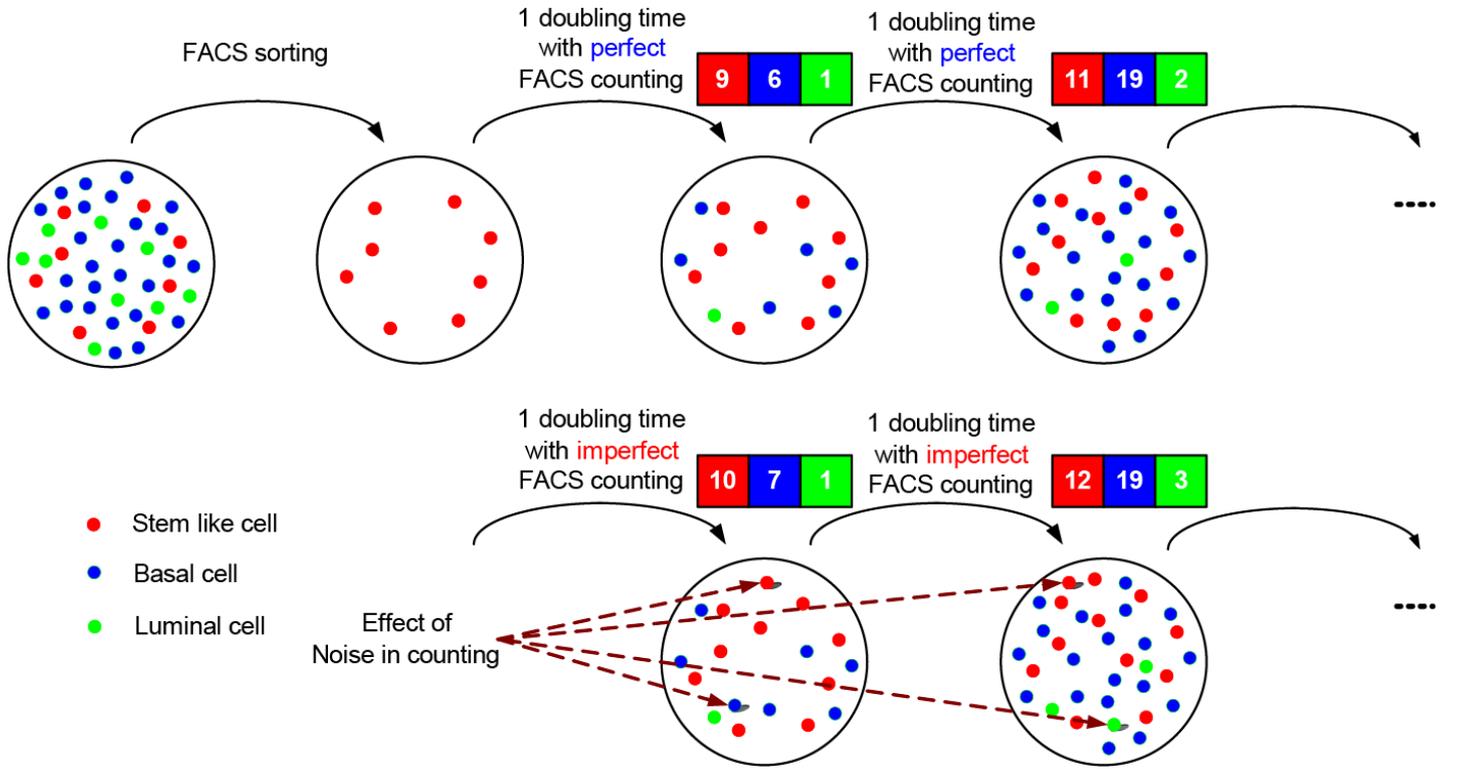

Fig. 2. The emergence of errors in counting breast cancer cells using FACS analysis (cf. [10])

After estimating the transition probability matrix using experimental data, we use that matrix for computer simulations of cell proliferation dynamics to support therapeutic decisions. Hence, in this stage another source of uncertainty in predicting the number of cells using the estimated transition probability matrix, is due to errors in estimating the state probability $\tilde{q}_j^{(k)}$, which is the probability of a cell being of type $j \in [1, M]$ at time $k$. The estimated state probability using LNN is obtained as

$$q_j^{(k)} = \frac{\tilde{v}_j^{(k)}}{\sum_{j=1}^{M} \tilde{v}_j^{(k)}}, \qquad (2)$$

where, $\tilde{v}_j^{(k)}$ is the population size of cell type $j$ at time $k$. The estimated state probability $q_j^{(k)}$ approaches $\tilde{q}_j^{(k)}$ for large $\tilde{v}_j^{(k)}$. Since only a noisy version (approximation) of population size of cell type $j$ at time $k$, $v_j^{(k)}$, is available, the estimated value for probability of cell type $j \in [1, M]$ at time $k$ is given by

$$\hat{q}_j^{(k)} = \frac{v_j^{(k)}}{\sum_{j=1}^{M} v_j^{(k)}}. \qquad (3)$$

In practice, *NMS* measurements are taken from each of the *NS* samples using FACS machine. Subsequently, using the proposed approach in this work, the transition probabilities are estimated. Next, one can use this to run Monte Carlo simulations to study cancer cell proliferation based on assessment of $\hat{q}_j^k$ over time, e.g., for drug design.

In summary, we consider the effects of two sources of uncertainty in predicting the number of different cancer cell types using an estimated transition probability matrix, $\hat{\mathbf{P}}$. The first source of uncertainty is the measurement noise, and the second source of uncertainty is the limitation in estimating $q_j^{(k)}$, here referred to as the randomness in sampling.

In practice, we cannot track the proliferation of a specific single cell over time but may observe portions of cells in different states. Fig. 3(a) shows portion of each type of cell at time $k$. Fig 3(b) shows a schematic diagram for Markov model of transition between cell portions. Fig. 3(c) shows schematic diagrams of hidden Markov model, when the size of cell population is measured in the presence of noise. In this model, the state probabilities are the alphabet of the hidden Markov model, which are continuous variables. In this model, the state probability is considered a random variable with Gaussian or Poisson distribution.

## 2.2. Problem Statement

Here, we formulate the effect of randomness in measurement and sampling on cell population counting. When the distribution of measurement noise is known, we use an MMSE or ML estimator to derive the transition probability matrix. For a formal definition of design optimization problems, we first define the related parameters in Table 2.

In FACS experiment analysis, the maximum number of measurements for cell population is limited. Because the cells die after a few time steps, and the dynamics of cell population growth change. Therefore, in our analysis we try to minimize the number of measurement in each sample. Moreover, the number of available samples is limited due to the cost of experiments; hence, we consider a limit on the maximum number of samples. Under these limitations, we estimate transition probability matrix defined in (1) from different samples. Our objective function is to minimize the number of measurement in each sample while ensuring that the transition probabilities are estimated to within a specified level of accuracy. Here, two performance measures are defined to compare these two cell populations in proposed optimization problems. First performance measure compares the cell populations of each cell types using *M* error functions for each cell type. In the second problem we average the error function of cell populations over different subpopulation sizes. This leads us to the following two design optimization problems. Note that to assess the performance of the proposed solutions for estimation of transition probability of the cell proliferation Markov process, we resort to simulations in Section 4. In this case, a true transition probability matrix is assumed and observations are obtained based on the model described.

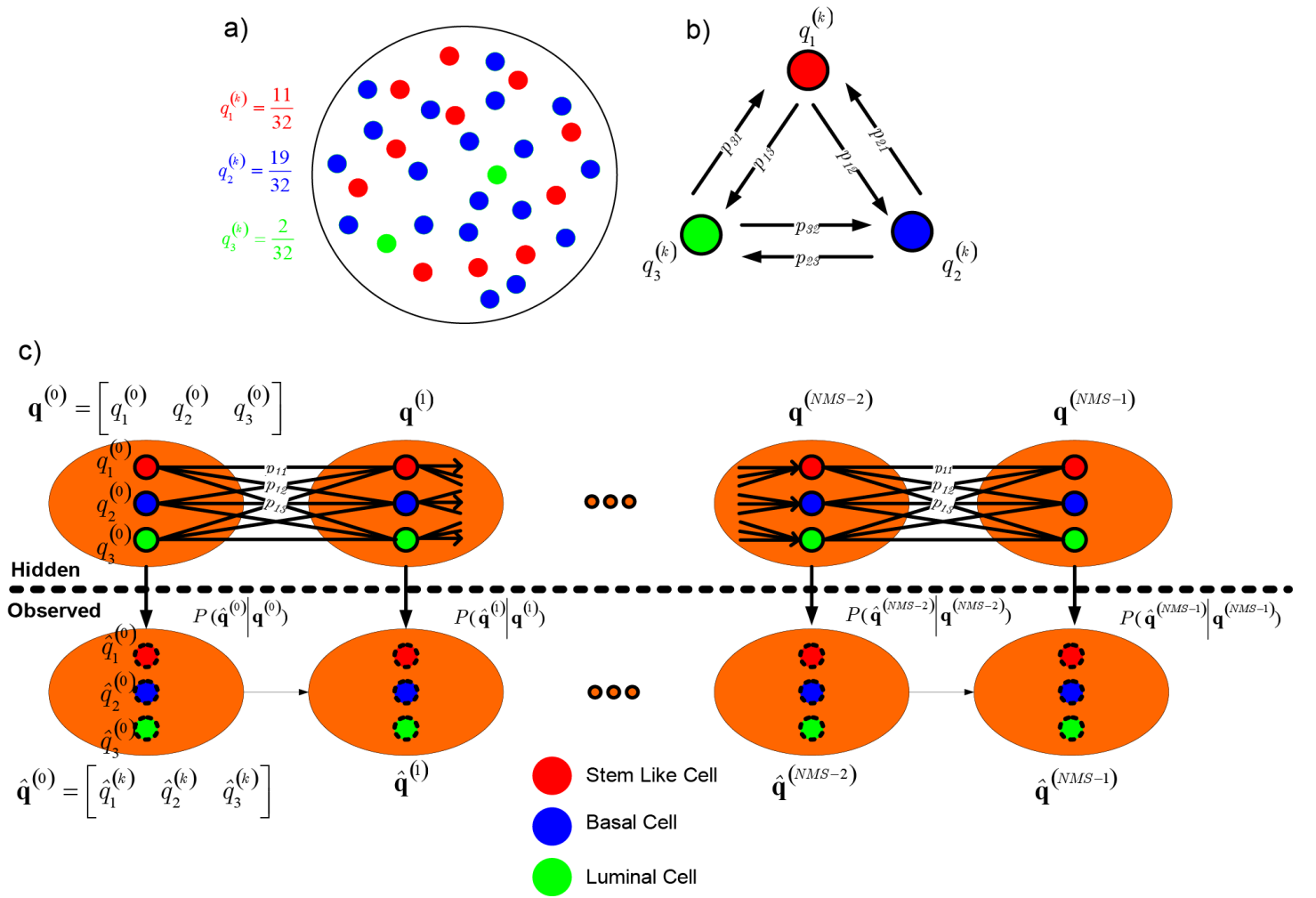

Fig. 3. Schematic diagram of state transitions in cell populations a) Portion of each type of cell at time $k$. In practice, we cannot measure single cell proliferation but can obtain the portion of each type of cell in a population. b) Schematic diagram of the Markov model describing transitions between portions of cells. c) Schematic diagram of a hidden Markov model, with noisy state probability measurements. In this case, the 'alphabet' $q^{(k)}$ of the model corresponds to state probabilities with continuous values.

Table. 2. Parameter definitions

| Sample | Sample means a preparation of cancer cells used for cell counting by FACS |
|---|---|
| Number of samples, $NS$ | The number of samples used for determining the transition probability matrix |
| Number of measurements in each sample, $NMS$ | The number of measurements in each tissue sample during the tumor progress |
| $PE_j^{(k)}$ | Percentage of error for subpopulation $j$ at time instant $k$ |
| $MPE^{(k)}$ | Mean of percentage of error over different subpopulations at time instant $k$ |
| $CV$ | Coefficient of variation (the ratio of standard deviation to mean) |

*Problem 1.* The minimum number of measurements required to limit the normalized error of predicted population size for each type of cell using the MMSE or ML estimated $\hat{\mathbf{P}}$ and limited $NS$, is computed in the following optimization problem

$$\min NMS$$
$$s.t.$$
$$PE_j^{(k)} \leq \varepsilon_j, j \in \{1,...,M\}$$
$$NS \leq NS_{max}, \qquad (4)$$

where $NS_{max}$ and $\varepsilon_j$ denote the maximum possible number of samples and the required accuracy in estimation of cell type $j$, respectively. Also, the percentage of error for subpopulation $j$ at time step $k$, $PE_j^{(k)}$ is given by

$$PE_j^{(k)} = \left| q_j^{(k)} - \hat{q}_j^{(k)} \right| \times 100. \quad (5)$$

In (4), $k$ indicates the (steady state) time instant in multiples of $T$ (the time step over which the number for cells is expected to double) at which the error is computed.

*Problem 2.* The minimum number of measurements required to limit the mean of normalized error of a predicted population size using the MMSE or ML estimated $\hat{\mathbf{P}}$ with limited $NS$, is computed in the following optimization problem

$$\begin{aligned}
&\min NMS \\
&\text{s.t.} \\
&\quad MPE^{(k)} \leq \varepsilon, \\
&\quad NS \leq NS_{max},
\end{aligned} \quad (6)$$

in which

$$MPE^{(k)} = \frac{1}{M} \sum_{j=1}^{M} PE_j^{(k)}, \quad (7)$$

and $\varepsilon$ denotes the average required accuracy in estimation of cell types. It can be verified that these problems are convex (details not reported here). In the next Section, we derive the MMSE estimator for $\hat{\mathbf{P}}$ when the cell population sizes are observed through additive Gaussian noise, which provides the same results as the ML estimation (see Supplementary Materials for details). Then, two ML estimators approximated based on one sample estimation and sample mean approximation are derived in the case of Poisson observations.

## 3. Stochastic Analysis

In the present Section, we provide results for the proposed MMSE and approximate ML estimators. We consider the case of noisy measurements, assuming a Gaussian distribution, which we found to fit with our experimental data. For small numbers of cells we assume a Poisson distribution, fitted to PDF of shot noise, to avoid consequences of the Gaussian distribution going into a range of negative values. The Gaussian distribution will allow us to derive a closed form solution. That is while for the Poisson distribution a closed form solution is not available, and we here provide an approximated solution for problems described in Propositions 1 and 2 and further discussed in the Supplementary Materials.

*3.1. Gaussian Distribution*

It is assumed that we have $N$ samples, where the initial cell population of each sample is unknown, and for sample $i$ the measurements of subpopulation sizes have a Gaussian distribution with mean $\begin{bmatrix} \tilde{v}_{1,i}^{(0)} & \tilde{v}_{2,i}^{(0)} & \cdots & \tilde{v}_{M,i}^{(0)} \end{bmatrix}$. The MMSE and ML estimators in a general form for $\hat{\mathbf{P}}$ are given by

$$\hat{\mathbf{P}}_{MMSE} = E\left(\mathbf{P} \big| v_{1,1}^{(0)}, ..., v_{j,i}^{(k)}, ..., v_{M,N}^{(NMS-1)}\right), \tag{8}$$

$$\hat{\mathbf{P}}_{ML} = \max_{\mathbf{P}} P\left(v_{1,1}^{(0)}, ..., v_{j,i}^{(k)}, ..., v_{M,N}^{(NMS-1)} \big| \mathbf{P}\right). \tag{9}$$

We assume for simplicity that the observed noise for different measurements are independent events. In general, the measurement noise may be due to physical and performance characteristics of the measurement equipment (here the FACS machine) and may be correlated. The exact study and characterization of such a possible correlated noise model is beyond the scope of the current work. Here, we derive the MMSE estimator when measurements of population sizes are corrupted with additive white Gaussian noise. Hence, due to Markov property for cell proliferation, the observation model is given by

$$2 \begin{bmatrix} \tilde{v}_{1,i}^{(0)} & \tilde{v}_{2,i}^{(0)} & \cdots & \tilde{v}_{M,i}^{(0)} \\ \tilde{v}_{1,i}^{(1)} & \tilde{v}_{2,i}^{(1)} & \cdots & \tilde{v}_{M,i}^{(1)} \\ \vdots & \vdots & \ddots & \vdots \\ \tilde{v}_{1,i}^{(NMS-2)} & \tilde{v}_{2,i}^{(NMS-2)} & \cdots & \tilde{v}_{M,i}^{(NMS-2)} \end{bmatrix} \begin{bmatrix} p_{1l} \\ p_{2l} \\ \vdots \\ p_{Ml} \end{bmatrix} + \begin{bmatrix} \eta_{l,i}^{(1)} \\ \eta_{l,i}^{(2)} \\ \vdots \\ \eta_{l,i}^{(NMS-1)} \end{bmatrix} = \begin{bmatrix} v_{l,i}^{(1)} \\ v_{l,i}^{(2)} \\ \vdots \\ v_{l,i}^{(NMS-1)} \end{bmatrix}, \tag{10}$$

where $\tilde{v}_{j,i}^{(k)}$ denotes exact values of cell population sizes. Also, $v_{l,i}^{(k)}$ and $\eta_{l,i}^{(k)}$ denote respectively, the measured values of cell population sizes and the noise term for cell type $l$, sample $i$ at time $k$. The noise term $\eta_{l,i}^{(k)}$ has a Gaussian distribution with zero mean and standard deviation $\sigma_l$. The factor 2 is due to the fact that the measurements take place at intervals when the cell population size is expected to double. In the following Theorem, the transition probability matrix of $\mathbf{P}$ is derived using an MMSE estimator.

***Theorem 1.*** The transition probability matrix, $\mathbf{P}$, of the cell proliferation Markov process, when the measurements of population sizes is observed in Gaussian noise, is obtained using an MMSE estimator as follows,

$$\mathbf{P}_{MMSE}^{T} = \begin{bmatrix} p_{11} & \cdots & p_{1M} \\ \vdots & \ddots & \vdots \\ p_{M1} & \cdots & p_{MM} \end{bmatrix}^{\dagger} =$$

$$\begin{bmatrix} \left( \left( \left(\mathbf{V}_{:,1:N}^{(0:NMS-2)}\right)^{\dagger} \left(\mathbf{V}_{:,1:N}^{(0:NMS-2)}\right) + N(NMS-1)\mathbf{R}_n \right)^{-1} \left(\mathbf{V}_{:,1:N}^{(0:NMS-2)}\right)^{\dagger} \mathbf{V}_{1,1:N}^{(1:NMS-1)} - \Lambda_1 \right)^{\dagger} \\ \vdots \\ \left( \left( \left(\mathbf{V}_{:,1:N}^{(0:NMS-2)}\right)^{\dagger} \left(\mathbf{V}_{:,1:N}^{(0:NMS-1)}\right) + N(NMS-1)\mathbf{R}_n \right)^{-1} \left(\mathbf{V}_{:,1:N}^{(0:NMS-2)}\right)^{\dagger} \mathbf{V}_{M,1:N}^{(1:NMS-1)} - \Lambda_M \right)^{\dagger} \end{bmatrix}, \quad (11)$$

where the superscript † denotes the transpose of the matrix and $\mathbf{V}_{:,1:N}^{(0:NMS-2)}$ is given by

$$\mathbf{V}_{:,1:N}^{(0:NMS-2)} = \begin{bmatrix} v_{1,1}^{(0)} & \cdots & v_{M,1}^{(0)} \\ \vdots & \ddots & \vdots \\ v_{1,1}^{(NMS-2)} & \cdots & v_{M,1}^{(NMS-2)} \\ \vdots & \vdots & \vdots \\ v_{1,N}^{(0)} & \cdots & v_{M,N}^{(0)} \\ \vdots & \ddots & \vdots \\ v_{1,N}^{(NMS-2)} & \cdots & v_{M,N}^{(NMS-2)} \end{bmatrix}, \quad (12)$$

also, $\mathbf{V}_{l,1:N}^{(1:M)}$, $\mathbf{R}_n$, is covariance matrix of noise, and $\Lambda_l$ is Lagrange multiplier, they are given by

$$\mathbf{V}_{l,(1:N)}^{(1:NMS-1)} = \begin{bmatrix} v_{l,1}^{(1)} & \cdots & v_{l,1}^{(M)} & \cdots & v_{l,N}^{(1)} & \cdots & v_{l,N}^{(NMS-1)} \end{bmatrix}^{\dagger} \quad (13)$$

$$\mathbf{R}_n = \begin{bmatrix} \sigma_1^2 & \cdots & 0 \\ \vdots & \ddots & \vdots \\ 0 & \cdots & \sigma_M^2 \end{bmatrix}, \quad (14)$$

$$\Lambda_l = \begin{bmatrix} \frac{\lambda_l}{4} & \cdots & \frac{\lambda_l}{4} \end{bmatrix}^T, \quad (15)$$

where $\sigma_l^2$ is noise variance of cell type $l$, $\lambda_l > 0$ is obtained by replacing $p_{jl}$ from (11) in the following equation for each value of $l$

$$\sum_{h=1}^{M} p_{hl} = 1. \quad (16)$$

The above solution is obtained when the constraints $\left(0 \leq p_{ij} \leq 1\right)$ is assumed satisfied. However, if the obtained results violate these constraints, one should enforce them and solve the problem again. The proof and further details are provided in Section B of the Supplementary Material.

This solution is also a solution of the ML estimator for $NMS-1 \geq M$ as elaborated in Section C of the Supplementary Material.

*B. Poisson Distribution*

Based on the distribution of shot noise, which is reported in [26-30], we here assume a Poisson distribution for cell population sizes measured in each sample $i \in \{1,...,N\}$:

$$\begin{bmatrix} v_{1,i}^{(0)} & \cdots & v_{M,i}^{(0)} \\ \vdots & \ddots & \vdots \\ v_{1,i}^{(NMS-1)} & \cdots & v_{M,i}^{(NMS-1)} \end{bmatrix} \sim \begin{bmatrix} \mathcal{P}\left(\tilde{v}_{1,i}^{(0)}\right) & \cdots & \mathcal{P}\left(\tilde{v}_{M,i}^{(0)}\right) \\ \vdots & \ddots & \vdots \\ \mathcal{P}\left(\tilde{v}_{1,i}^{(NMS-1)}\right) & \cdots & \mathcal{P}\left(\tilde{v}_{M,i}^{(NMS-1)}\right) \end{bmatrix}. \tag{17}$$

Here, the observation model, which can be obtained from the ML estimator (See Section D of Supplementary Materials), is given by

$$2 \begin{bmatrix} \tilde{v}_{1,i}^{(0)} & \tilde{v}_{2,i}^{(0)} & \cdots & \tilde{v}_{M,i}^{(0)} \\ \tilde{v}_{1,i}^{(1)} & \tilde{v}_{2,i}^{(1)} & \cdots & \tilde{v}_{M,i}^{(1)} \\ \vdots & \vdots & \ddots & \vdots \\ \tilde{v}_{1,i}^{(NMS-2)} & \tilde{v}_{2,i}^{(NMS-2)} & \cdots & \tilde{v}_{M,i}^{(NMS-2)} \end{bmatrix} \begin{bmatrix} p_{1l} \\ p_{2l} \\ \vdots \\ p_{Ml} \end{bmatrix} = \begin{bmatrix} v_{l,i}^{(1)} \\ v_{l,i}^{(2)} \\ \vdots \\ v_{l,i}^{(NMS-1)} \end{bmatrix}, \tag{18}$$

In this setting, the observation noise is signal dependent and not additive. As a result, obtaining a closed form MMSE solution is challenging even in the case of independent Poisson observations. We can however compute two approximate ML solutions for the problem with Poisson observations in the next following Propositions.

**Proposition 1.** The transition probability matrix, $\mathbf{P}$, for the cell proliferation Markov process, when the measurements of population sizes are Poisson distributed, is obtained using an approximate ML one sample estimator as follows:

$$\mathbf{P}_{ML}^T = 0.5 \begin{bmatrix} 0.5\left(\left(\mathbf{V}_{:,1:N}^{(0:M-1)}\right)^\dagger \left(\mathbf{V}_{:,1:N}^{(0:M-1)}\right)\right)^{-1} \left(\mathbf{V}_{:,1:N}^{(0:M-1)}\right)^\dagger \mathbf{V}_{1,(1:N)}^{(1:M)} - \Lambda_1, \\ \vdots \\ 0.5\left(\left(\mathbf{V}_{:,1:N}^{(0:M-1)}\right)^\dagger \left(\mathbf{V}_{:,1:N}^{(0:M-1)}\right)\right)^{-1} \left(\mathbf{V}_{:,1:N}^{(0:M-1)}\right)^\dagger \mathbf{V}_{M,(1:N)}^{(1:M)} - \Lambda_M, \end{bmatrix} \tag{19}$$

where $\Lambda_l$ is Lagrange multiplier and is given by

$$\Lambda_l = \begin{bmatrix} \dfrac{\lambda_l}{4} & \cdots & \dfrac{\lambda_l}{4} \end{bmatrix}^\dagger, \tag{20}$$

and $\lambda_l > 0$ is obtained by solving the following equation for each value of $l$

$$\sum_{j=1}^{M} p_{jl} = 1. \tag{21}$$

The above solution is obtained when the constraints $\left(0 \leq p_{ij} \leq 1\right)$ is assumed satisfied. However, if the obtained results violate these constraints, one should enforce them and solve the problem again. See Section E of the Supplementary Materials for details and proof.

**Proposition 2.** The transition probability matrix, $\mathbf{P}$, for the cell proliferation Markov process, when the measurements of population sizes are Poisson distributed, is obtained using an ML estimator based on approximated sample mean as follows,

$$\hat{\mathbf{P}}_{ML}^{T} = 0.5 \begin{bmatrix} \left(\left(\boldsymbol{\Sigma}_v^\dagger \boldsymbol{\Sigma}_v\right)^{-1} \boldsymbol{\Sigma}_v^\dagger \boldsymbol{\Psi}_1 - \Lambda_1\right)^\dagger \\ \vdots \\ \left(\left(\boldsymbol{\Sigma}_v^\dagger \boldsymbol{\Sigma}_v\right)^{-1} \boldsymbol{\Sigma}_v^\dagger \boldsymbol{\Psi}_M - \Lambda_M\right)^\dagger \end{bmatrix}, \tag{22}$$

where $\Lambda_l$ is given by

$$\Lambda_l = \begin{bmatrix} \dfrac{\lambda_l}{4} & \cdots & \dfrac{\lambda_l}{4} \end{bmatrix}^\dagger, \tag{23}$$

and $\lambda_l > 0$ is obtained from the following equation for each value of $l$

$$\sum_{j=1}^{M} p_{jl} = 1, \tag{24}$$

and $\boldsymbol{\Sigma}_v$ is given by

$$\boldsymbol{\Sigma}_v = \begin{bmatrix} \sum_{i=1}^{N} v_{1,i}^{(0)} & \cdots & \sum_{i=1}^{N} v_{M,i}^{(0)} \\ \vdots & \ddots & \vdots \\ \sum_{i=1}^{N} v_{1,i}^{(NMS-2)} & \cdots & \sum_{i=1}^{N} v_{M,i}^{(NMS-2)} \end{bmatrix}. \tag{25}$$

Also, $\Psi_l$ is given by

$$\Psi_l = \left[ \sum_{i=1}^{N} v_{l,i}^{(1)} \quad \cdots \quad \sum_{i=1}^{N} v_{l,i}^{(NMS-1)} \right]. \tag{26}$$

The above solution is obtained when the constraints $\left(0 \leq p_{ij} \leq 1\right)$ is assumed satisfied. However, if the obtained results violate these constraints, one should enforce them and solve the problem again. See Section F of the Supplementary Materials for details and proof.

## 4. Simulation Results

In this Section, we provide simulation and numerical results for the optimization problems 1 and 2. Moreover, the dynamics of the convergence of cell population sizes are studied using a Monte Carlo simulation.

### 4.1. Simulation Scenario

For simulations, the number of cells in each sample is considered a random variable with uniform distribution between 3000 and 6000 cells. The number of measurements in each sample, $NMS$, is selected a multiple of $M$, where in line with [10], the number of states, $M$, is set to three, which corresponds to the three cell types of stem-like cells, basal cells and luminal cells. As stated, two scenarios with either additive white Gaussian observation noise or with Poisson observations are considered. The transition probability matrix for the cell proliferation Markov process simulated in this Section is selected as in [10]. This is also reported in the first row of Table 3.

### 4.2. Numerical Results

In this Section, first the MMSE estimator of Theorem 1 is used for the Gaussian scenario and the approximate ML estimator of Proposition 1 for the Poisson scenario. Our experiments reveal similar performance for the approximate ML estimators in Propositions 1 and 2. The performance of the proposed estimators are quantified by comparing the obtained results with the true (postulated) transition probability in the simulations. Second, using Monte Carlo computer simulations for each cell, we assess the cell proliferation by obtaining the PDF of $\tilde{q}_j^{(i)}$, $j \in \{1, 2, 3\}$, over different time steps $k \in \{T, 2T, \ldots, 10T\}$.

Fig. 4(a) shows the percentage of error in estimation of stem-like cell population, $PE_1^{(k)}$, in terms of the time step index, $k$, for different number of samples, $NS$, and number of measurements in each sample, $NMS$, set to 3. The measurement noise is Gaussian with coefficient of variation (CV) of 0.2236, or alternatively power of signal to power of noise ratio (SNR) of 13 dB; Based on the experimental data reported in part in Fig. 1, a SNR between 10 to 15 dB is considered typical. Figs. 4(b), (c) and (d) depict the results of similar experiments, but with *NMS* set to 6, 9 and 12, respectively. The results demonstrate that as expected $PE_1^{(k)}$ reduces as the number of samples, *NS*, or the number of measurements in each sample, *NMS*, increases.

Fig. 5(a) shows the percentage of error for estimation of stem like cell population in steady state ($PE_1^{(k)}$ for $k = 20$), in terms of *NS* and *NMS* for additive white Gaussian measurement noise. Figs. 5(b) and 5(c) shows the results of similar experiments for the case of basal and luminal cell subpopulations, respectively. The mean percentage of error in steady state ($k = 20$) estimation of cell population sizes, *MPE*, is depicted in Fig. 5(d). Fig. 5 shows that if the number of measurements in each sample is small, increasing the number of samples, *NS*, generally improves the accuracy of estimation, but it also shows that there are fluctuations in the accuracy of estimation. By increasing the number of measurements in each sample, *NMS*, the fluctuations disappear. This means that for given *NS*, we have a lower bound on *NMS*. For example, if we have 10 samples, $NS = 10$ and in each sample we only have one measurement, the accuracy of estimation is therefore not guaranteed. A tradeoff exists then between *NS* and *NMS*. Increasing *NS* increases the costs for experiments, this means that we need more samples for measurements. Moreover, increasing the number of measurements for cell populations is usually limited for practical purposes because after some steps the cells die and the dynamics of cell population growth change. As a result, Figs. 5(a)-(c) provide a numerical solution for the optimization problem 1; For example for $j = 1$, setting $NS = NMS = 6$, limits the *PE* under 5%. A conclusion is thus that one should select the best values for *NS* and *NMS* to satisfy the related accuracy and practical constraints.

Figs. 6 (a)-(c) show the percentage of estimation error, *PE*, for the proposed approximate ML estimator in Proposition 1, in terms of *NS* and *NMS* with measurement shot noise for respectively the stem-like, the basal, and the luminal cells, respectively. Fig. 6(d) shows the mean percentage of error for the same experiment. As evident increasing *NS* and *NMS*, reduces *PE* and for $NS = 4$ and $NMS = 3$, it is less than 1% for all cell subpopulations in cell proliferation process. The Figs. 6(a)-(c) thus provide numerical solutions for the optimization Problem 1.

Table 3. Estimated transition probability matrix obtained with different approaches with Gaussian noise ($CV = 0.2236$).

| Exact transition probability matrix | $P = \begin{bmatrix} 0.58 & 0.35 & 0.07 \\ 0.01 & 0.99 & 0 \\ 0.04 & 0.49 & 0.47 \end{bmatrix}$ |
| --- | --- |
| Estimated transition probability matrix without measurement and sampling randomness | $\hat{P} = P$ |
| Estimated transition probability matrix in presence of noise but ignoring it in the estimation | $\hat{P} = \begin{bmatrix} 0.0743 & 0.9257 & 0 \\ 0 & 0.0167 & 0.9833 \\ 0.0659 & 0 & 0.9341 \end{bmatrix}$ |
| Estimated transition probability matrix based on MMSE in presence of noise; $NMS = 12$ and $NS = 6$. | $\hat{P} = \begin{bmatrix} 0.5614 & 0.3590 & 0.0795 \\ 0.0203 & 0.9764 & 0.0033 \\ 0.0293 & 0.5149 & 0.4558 \end{bmatrix}$ |

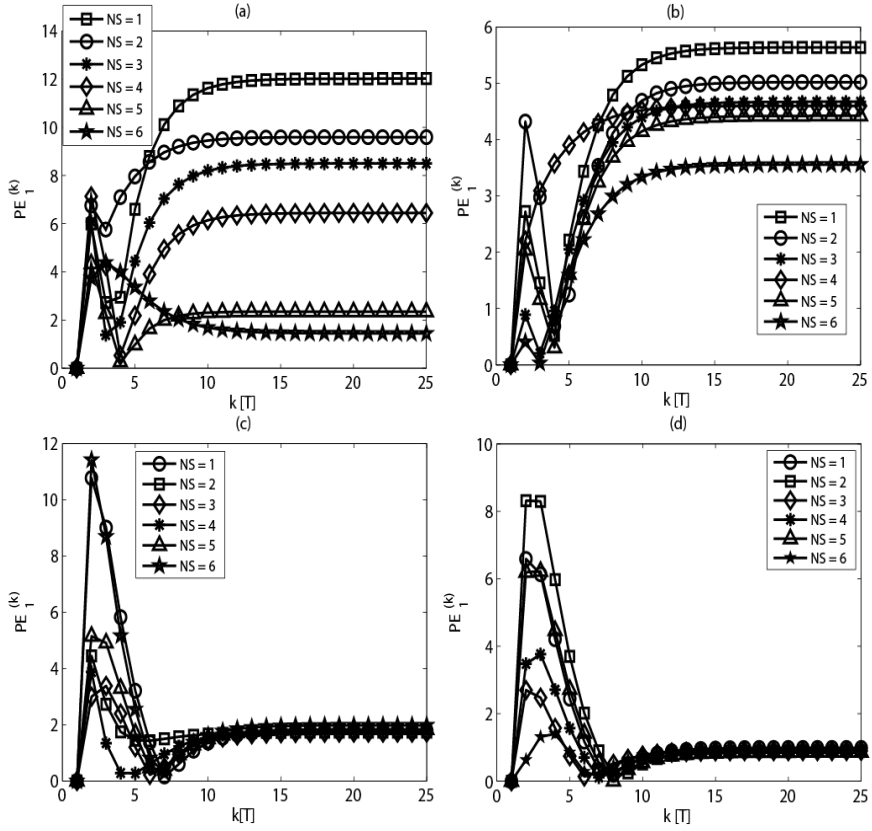

Fig. 4. $PE_1^{(k)}$ in presence of Gaussian measurement noise of $CV = 0.2236$, as a function of $k$ for different values of $NS$ and with (a) $NMS = 3$, (b) $NMS = 6$, (c) $NMS = 9$, and (d) $NMS = 12$.

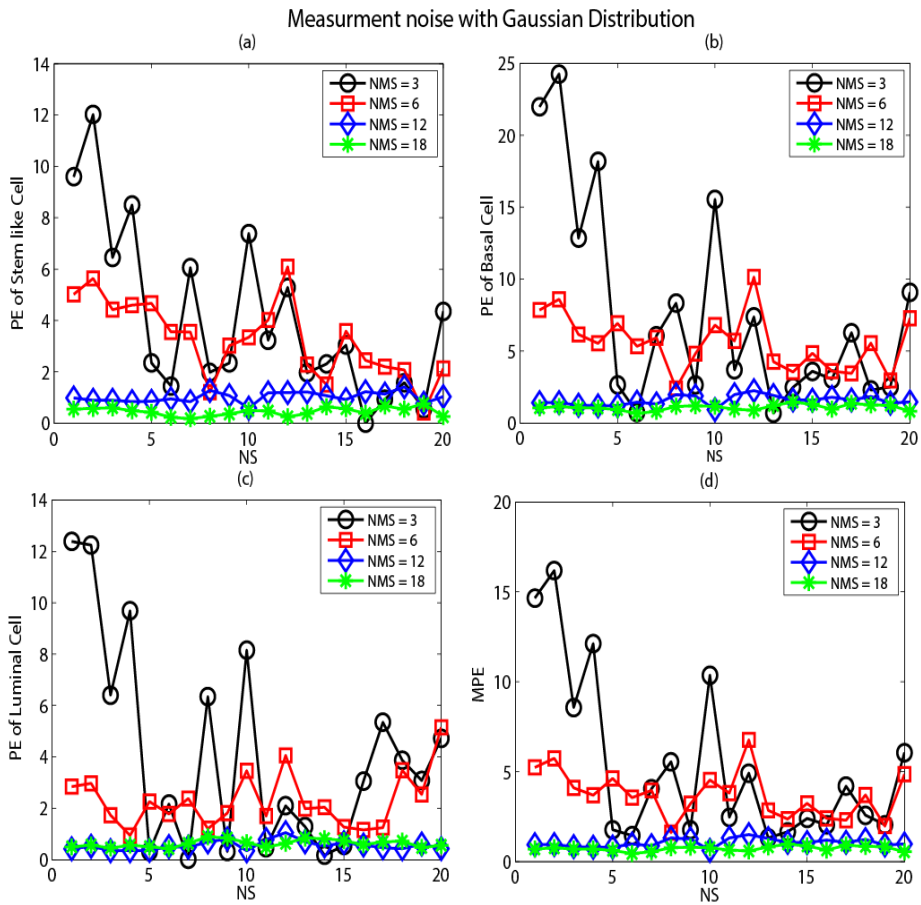

Fig. 5. *PE* for the proposed MMSE estimator in presence of Gaussian measurement noise of *CV* = 0.2236, as a function of *NS* for different values of *NMS* (a) Stem-like cell (b) Basal cell (c) Luminal cell. (d) *MPE*

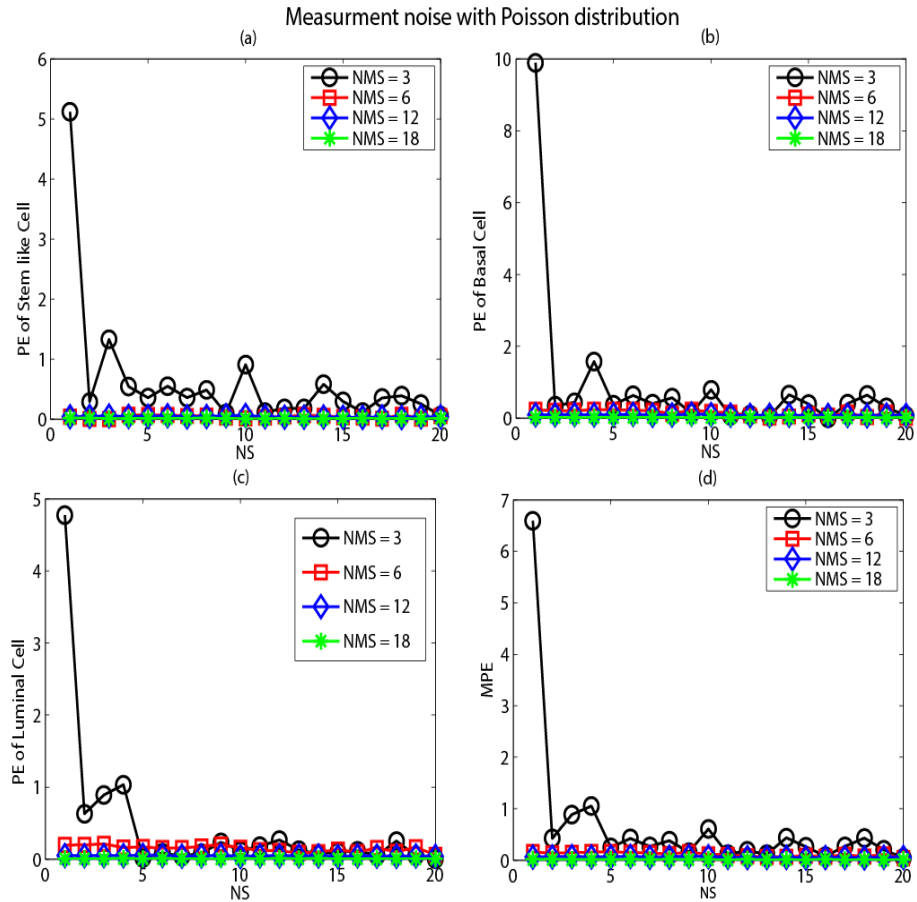

Fig. 6. *PE* for the proposed approximate ML estimator in presence of shot noise (Poisson observations), as a function of *NS* for different values of *NMS* (a) Stem-like cell (b) Basal cell (c) Luminal cell. (d) *MPE*

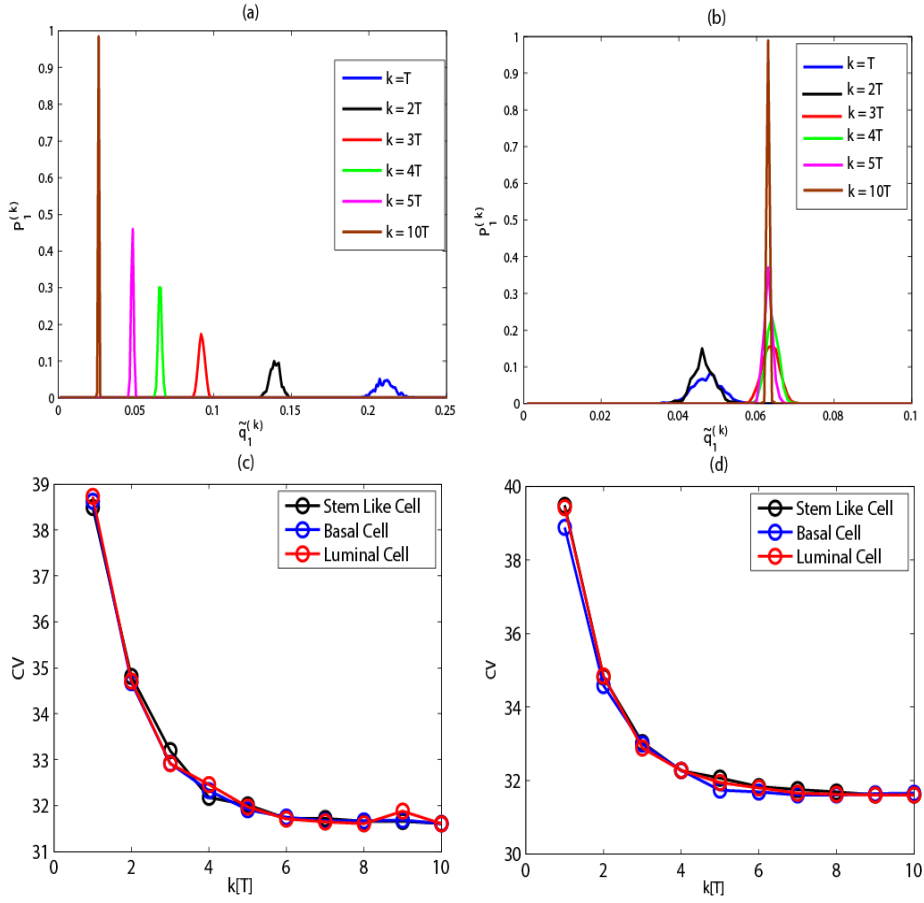

Fig. 7(a). PDF of $\tilde{q}_S^{(k)}$ i.e. cell state probability of stem like cell for different values of $k$, when $\hat{\mathbf{P}}$ is derived using the MMSE estimator. (b) using a deterministic approach. (c) $CV$ in terms of $k$, for stem-like, basal and luminal cell populations when $P$ is derived using MMSE estimator, (d) using a deterministic approach.

Figs. 5(d) and 6(d) provide numerical solutions for optimization Problem 2 and measurement Gaussian and shot noises, respectively. It is evident in Fig. 5(d) that $MPE$ is bound to 5% for $NS = NMS = 6$. Fig. 6(d) shows $MPE$ is less than 1% for $NS = 5$ and $NMS = 3$.

Fig. 7(a) shows the PDF of $\tilde{q}_1^{(k)}$, i.e., state probability of stem like cell, which is denoted by $P_1^{(k)}$ for different time steps of $k$, when the transition probability matrix is derived using MMSE estimator from noisy data with Gaussian distribution, and with the transition probability matrix shown in Table 3. This PDF is observed by Markov chain Monte Carlo simulation. It can be seen that by increasing $k$, the variance of $\tilde{q}_S^{(k)}$ is reduced and converges to a steady state value. Fig. 7(b) shows $P_1^{(k)}$ for different values of $k$, when the transition probability matrix is derived using the deterministic approach (see Table 3). It can be clearly seen, that the dynamics of convergence for the Markov process and the steady state value of $\tilde{q}_1^{(k)}$ are distinct. Figs. 7 (c) and (d) show, the coefficient of variation ($CV$) of $P_j^{(k)}$ in terms of $k$, which is defined by the ratio of standard deviation to mean of $\tilde{q}_1^{(k)}$ when transition probability matrix is obtained based on the MMSE estimator and the deterministic approach. In both cases, the *CV* is constant after seven doubling steps, which shows that dispersion of $P_j^{(i)}$ during time is independent of estimation method of transition probability matrix.

## 5. Conclusions

Predicting the temporal evolution of cell population sizes using Florescent Activated Cell Sorting (FACS) is an important task for a variety of biological and biomedical applications. A specific example, which has motivated our research, comes from cancer research where the dynamics of population sizes for normal and cancer stem cells can provide clues for therapeutic decisions. Studying proliferating cell populations and their proportions in tissue samples, Markov processes provide a suitable conceptual framework to model and simulate such systems. A Markov process is characterized by a transition matrix that assembles the probabilities for transitions between states (here cell types). In the present paper, the transition matrix of the Markov process is estimated from aggregated data of FACS measurements. In this context, noisy measurements are used for the estimation of the transition probability matrix. Sampling randomness, on the other hand, is here related to the error in estimating the state probability from small cell populations. If noiseless observations were available, only $M^2$ measurements would be required for estimating the transition probability matrix; but this situation does not reflect the situation in experimental laboratories. Assuming the realistic scenario of noisy data, an exact prediction of population sizes directly depends on the number of samples analyzed. In our first example, the sample and $M^2$ measurements for population sizes during cell doubling were enough to estimate the transition probability matrix perfectly, allowing exact prediction of subpopulation sizes. We then showed that if measurements are noisy, the number of samples to be analyzed and the number of measurement for each sample must be carefully considered. We subsequently developed a MMSE estimator for calculating the transition probability matrix when counts of the cell population are corrupted by Gaussian noise. Moreover, in the case of Poisson noise, we derive two approximate ML solution using one sample estimator and sample mean approximation. Our numerical results show that, if the deterministic approach is used for estimating the transition probability matrix, the prediction of subpopulation sizes can easily be erroneous. We demonstrate, that the prediction of the convergence for the proliferation process and the resulting steady states can substantially differ, which would have obvious consequences for the design of therapies that rely on this information.

An interesting point, arising from our simulation study, is that the coefficient of variation of the PDF for the cell population probability is not changed during proliferation as a consequence of noise in measurements. Numerical results show, that, when there is shot noise and with the number of samples *NS*=5 and the number of measurements per sample, *NMS*=3, the mean of error in predicting the population size (*MPE*) is less than 1%.

Moreover, for noise with a Gaussian distribution and coefficient of variation of 0.2236, the mean of error in predicting population size $MPE$ was less 5%, when $NS = NMS = 6$ is considered. Our results show that increasing $NMS$ is more effective than increasing $NS$ in improving accuracy of transition probability matrix estimation. Our study also showed the consequences of estimating the probability of a cell being in a particular state from measurements across a population of cells. For small population sizes, the law of large numbers will not be satisfied, leading to errors. We showed that the uncertainty arising from small sample sizes can be leveled by using a distribution for the state probability rather than a single value. Our work thus contributes to a better understanding of randomness and noise when studying stochastic phenomena inevitably linked to FACS experiments.

## 6. Acknowledgments

This research has been supported in part by the DAAD (German-Arabic/Iranian Higher Education Dialogue).

## 7. References

1       Medema, J.P.: 'Cancer stem cells: the challenges ahead', Nature cell biology, 2013, 15, (4), pp. 338-344
2       Vermeulen, L., Sprick, M.R., Kemper, K., Stassi, G., and Medema, J.P.: 'Cancer stem cells - old concepts, new insights', Cell Death Differ, 2008, 15, (6), pp. 947-958
3       Kemper, K., Sprick, M.R., de Bree, M., Scopelliti, A., Vermeulen, L., Hoek, M., Zeilstra, J., Pals, S.T., Mehmet, H., Stassi, G., and Medema, J.P.: 'The AC133 Epitope, but not the CD133 Protein, Is Lost upon Cancer Stem Cell Differentiation', Cancer Res, 2010, 70, (2), pp. 719-729
4       Mak, A.B., Blakely, K.M., Williams, R.A., Penttila, P.A., Shukalyuk, A.I., Osman, K.T., Kasimer, D., Ketela, T., and Moffat, J.: 'CD133 Protein N-Glycosylation Processing Contributes to Cell Surface Recognition of the Primitive Cell Marker AC133 Epitope', J Biol Chem, 2011, 286, (47), pp. 41046-41056
5       La Porta, C.A.M., Zapperi, S., and Sethna, J.P.: 'Senescent Cells in Growing Tumors: Population Dynamics and Cancer Stem Cells', Plos Comput Biol, 2012, 8, (1)
6       Lapidot, T., Sirard, C., Vormoor, J., Murdoch, B., Hoang, T., Caceresсortes, J., Minden, M., Paterson, B., Caligiuri, M.A., and Dick, J.E.: 'A Cell Initiating Human Acute Myeloid-Leukemia after Transplantation into Scid Mice', Nature, 1994, 367, (6464), pp. 645-648
7       Visvader, J.E.: 'Cells of origin in cancer', Nature, 2011, 469, (7330), pp. 314-322
8       Vermeulen, L., Melo, F.D.E., Richel, D.J., and Medema, J.P.: 'The developing cancer stem-cell model: clinical challenges and opportunities', Lancet Oncol, 2012, 13, (2), pp. E83-E89
9       Magee, J.A., Piskounova, E., and Morrison, S.J.: 'Cancer Stem Cells: Impact, Heterogeneity, and Uncertainty', Cancer Cell, 2012, 21, (3), pp. 283-296
10      Gupta, P.B., Fillmore, C.M., Jiang, G.Z., Shapira, S.D., Tao, K., Kuperwasser, C., and Lander, E.S.: 'Stochastic State Transitions Give Rise to Phenotypic Equilibrium in Populations of Cancer Cells (vol 146, pg 633, 2011)', Cell, 2011, 147, (5), pp. 1197-1197
11      Zapperi, S., and La Porta, C.A.M.: 'Do cancer cells undergo phenotypic switching? The case for imperfect cancer stem cell markers', Sci Rep-Uk, 2012, 2
12      Ullah, M., and Wolkenhauer, O.: 'Stochastic approaches in systems biology', Wires Syst Biol Med, 2010, 2, (4), pp. 385-397
13      Wallace, E.W.J., Gillespie, D.T., Sanft, K.R., and Petzold, L.R.: 'Linear noise approximation is valid over limited times for any chemical system that is sufficiently large', Iet Syst Biol, 2012, 6, (4), pp. 102-115
14      Thomas, P., Straube, A.V., and Grima, R.: 'The slow-scale linear noise approximation: an accurate, reduced stochastic description of biochemical networks under timescale separation conditions', Bmc Syst Biol, 2012, 6
15      Roussel, M.R., and Tang, T.: 'Simulation of mRNA diffusion in the nuclear environment', Iet Syst Biol, 2012, 6, (4), pp. 125-U162
16      Grima, R.: 'Intrinsic biochemical noise in crowded intracellular conditions', J Chem Phys, 2010, 132, (18)
17      Greese, B., Wester, K., Bensch, R., Ronneberger, O., Timmer, J., Hulskamp, M., and Fleck, C.: 'Influence of cell-to-cell variability on spatial pattern formation', Iet Syst Biol, 2012, 6, (4), pp. 143-153 18     Herzenberg, L.A., Tung, J., Moore, W.A., Herzenberg, L.A., and Parks, D.R.: 'Interpreting flow cytometry data: a guide for the perplexed', Nat Immunol, 2006, 7, (7), pp. 681-685
19     Yen-Chen Fu, A.: 'Microfabricated Fluorescence-Activated Cell Sorters (mu FACS) for Screening Bacterial Cells', California Institute of Technology, 2002
20     Kalbfleisch, J.D., Lawless, J.F., and Vollmer, W.M.: 'Estimation in Markov-Models from Aggregate Data', Biometrics, 1983, 39, (4), pp. 907-919
21     Kalbfleisch, J.D., and Lawless, J.F.: 'Least-Squares Estimation of Transition Probabilities from Aggregate Data', The Canadian Journal of Statistics, 1984, 12, (3), pp. 169-182
22     Baum, L.E., Petrie, T., Soules, G., and Weiss, N.: 'A Maximization Technique Occurring in Statistical Analysis of Probabilistic Functions of Markov Chains', Ann Math Stat, 1970, 41, (1), pp. 164-&
23     Maletzki, C., Stier, S., Gruenert, U., Gock, M., Ostwald, C., Prall, F., and Linnebacher, M.: 'Establishment, Characterization and Chemosensitivity of Three Mismatch Repair Deficient Cell Lines from Sporadic and Inherited Colorectal Carcinomas', PLoS One, 2012, 7, (12)
24     Tour, O., Adams, S.R., Kerr, R.A., Meijer, R.M., Sejnowski, T.J., Tsien, R.W., and Tsien, R.Y.: 'Calcium Green FlAsH as a genetically targeted small-molecule calcium indicator', Nat Chem Biol, 2007, 3, (7), pp. 423-431
25     Blanter, Y.M., and Buttiker, M.: 'Shot noise in mesoscopic conductors', Phys Rep, 2000, 336, (1-2), pp. 1-166
26     Stavis, S.M., Edel, J.B., Li, Y., Samiee, K.T., Luo, D., and Craighead, H.G.: 'Detection and identification of nucleic acid engineered fluorescent labels in submicrometre fluidic channels', Nanotechnology, 2005, 16, (7), pp. S314-323
27     Anderson, M.T., Tjioe, I.M., Lorincz, M.C., Parks, D.R., Herzenberg, L.A., Nolan, G.P., and Herzenberg, L.A.: 'Simultaneous fluorescence-activated cell sorter analysis of two distinct transcriptional elements within a single cell using engineered green fluorescent proteins', Proceedings of the National Academy of Sciences of the United States of America, 1996, 93, (16), pp. 8508-8511
28     Huebner, A., Srisa-Art, M., Holt, D., Abell, C., Hollfelder, F., deMello, A.J., and Edel, J.B.: 'Quantitative detection of protein expression in single cells using droplet microfluidics', Chemical communications, 2007, (12), pp. 1218-1220
29     Hill, E.K., and de Mello, A.J.: 'Single-molecule detection using confocal fluorescence detection: Assessment of optical probe volumes', Analyst, 2000, 125, (6), pp. 1033-1036
30     Adams, D.S., and Levin, M.: 'General principles for measuring resting membrane potential and ion concentration using fluorescent bioelectricity reporters', Cold Spring Harbor protocols, 2012, 2012, (4), pp. 385-397
31     Fillmore, C.M., and Kuperwasser, C.: 'Human breast cancer cell lines contain stem-like cells that self-renew, give rise to phenotypically diverse progeny and survive chemotherapy', Breast Cancer Res, 2008, 10, (2)
32     Werner, J.H., McCarney, E.R., Keller, R.A., Plaxco, K.W., and Goodwin, P.M.: 'Increasing the resolution of single pair fluorescence resonance energy transfer measurements in solution via molecular cytometry', Anal Chem, 2007, 79, (9), pp. 3509-3513
33     Tkacik, G.: 'From statistical mechanics to information theory: understanding biophysical information-processing systems', in Editor (Ed.)^(Eds.): 'Book From statistical mechanics to information theory: understanding biophysical information-processing systems' (2010, edn.), pp. 1-52

# Supplementary Materials: Accounting for Randomness in Measurement and Sampling in Study of Cancer Cell Population Dynamics

Siavash Ghavami[1*,**], Olaf Wolkenhauer[2**,***], Farshad Lahouti[3*], Mukhtar Ullah[4**], Michael Linnebacher[5****]

**Section A: Flow cytometric analysis of HROC87**

Flow cytometric analysis of HROC87, a cell line recently established from a primary colorectal cancer [1]. Exemplary data from 1 of 50 (S1) measurements and from 1 of 100 (S2) measurements are given. Similar data have also been generated from another cell line: HROC113 (data not shown). Of note, the analysis was performed in very low passages of the cell line.

The cells were harvested from cell culture in the exponential growth phase (approximately 80% density), washed with phosphate-buffered saline and incubated with 5µM Vybrant®Dye Cycle$^{TM}$ Violet Stain (VDC; Life Technologies, Frankfurt, Germany) in hanks-balanced salt solution for 30min at 37°C in the dark. In the control measurements (S1), 50µM Verapamil (Sigma-Aldrich, Hamburg, Germany) was added before the addition of VDC to block the dye-efflux by membrane-bound pumps. Cells were kept at 37°C until analysis for a maximum of 3 hours. Propidium iodide (1µg/ml; Life Technologies) was added shortly before measurement to allow for life/dead cell discrimination.

Samples were analyzed on a FACS ARIA II cell sorter equipped with standard lasers using the Diva software package (both from Becton Dickinson, Heidelberg, Germany). 100.000 events (all dots in S1(a) and S2(a) were counted per measurement.

Emails: [1]s.ghavami@ut.ac.ir, [2]olaf.wolkenhauer@uni-rostock.de, [3]lahouti@ut.ac.ir, [4]mukhtar.ullah@nu.edu.pk, [5]michael.linnebacher@med.uni-rostock.de
* Center for Wireless Multimedia Communications, University of Tehran, Tehran, Iran, wmc.ut.ac.ir
** Department of Systems Biology and Bioinformatics, University of Rostock, Rostock, Germany, www.sbi.uni-rostock.de.
*** Stellenbosch Institute for Advanced Study (STIAS), Wallenberg Research Centre at Stellenbosch University, Stellenbosch, South Africa.
**** Department of General, Thoracic, Vascular and Transplantation Surgery, University of Rostock, Rostock, Germany, http://www.moi.med.uni-rostock.de/

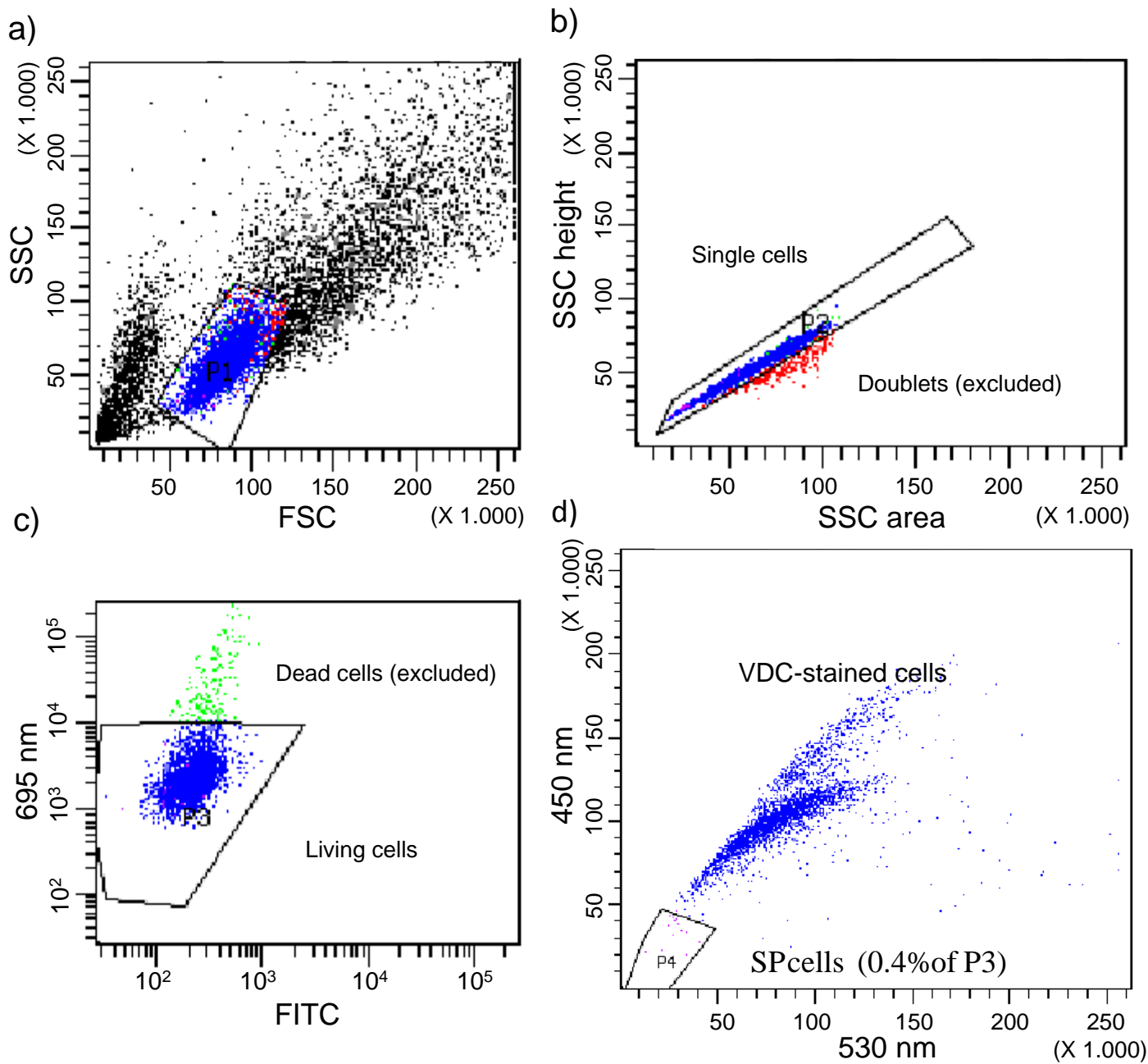

Fig. S1. Exemplary data from 1 of 50 measurements a) Cell size and granularity b) Exclusion of doublets c) Exclusion of dead cells $P_3$ d) SP Determination.

The gating strategy was as follows. S1(a) and S2(a): gate P1 was set in the forward scatter (FSC; cell size;X-axis) / sideward scatter (SSC; cell granularity;Y-axis) blot on the main cell population (blue dots). S1(b) and S2(b): In the SSC-area (X-axis) versus SSC-height blot (Y-axis), gate P2 was set to exclude doublets (red dots). S1(c) and S2(c): Dead cells were excluded by gating on the PI negative cells measured in the 695 channel (Y-axis; P3; green dots). For a better display, the blots additionally give the empty Fluorescein isothiocyanate (FITC) channel (X-axis). S1(d) and S2(d). Finally, the events were displayed in the 530 nm channel (X-axis) versus the V450 channel (Y-axis). The side population (SP) cells are those able to pump the VDC stain out of their cytoplasm (pink dots). They are not positively stained for VDC and lie within gate P4. Percentages of cells within P4 are given for the control cells (incubated with Verapamil and VDC; S1 and for the SP cell analysis (incubated with VDC but without inhibitor; S2.)

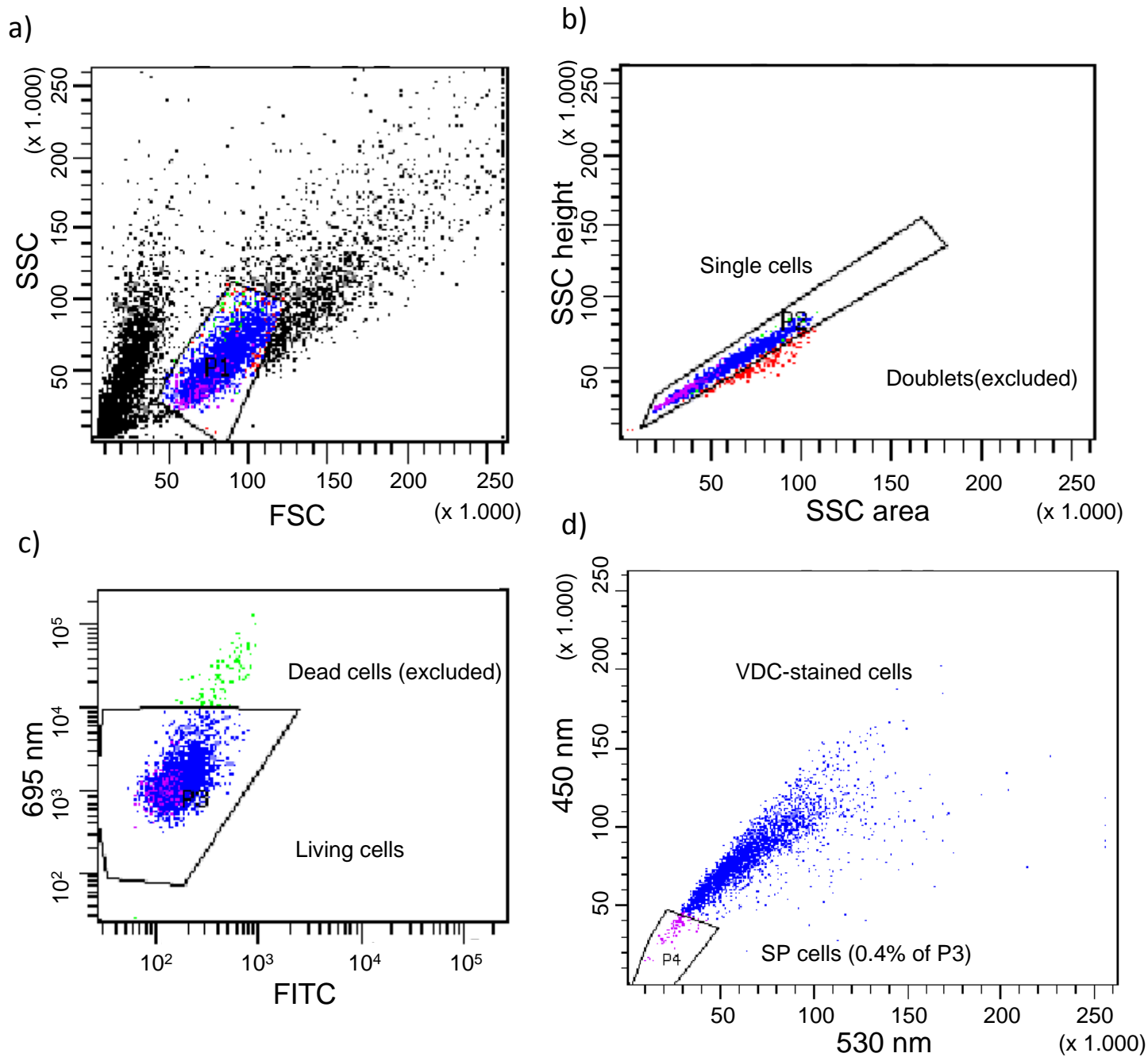

Fig. S2. Exemplary data from 1 of 100 measurements a) Cell size and granularity b) Exclusion of doublets c) Exclusion of dead cells $P_3$ d) SP Determination.

**Section B. Proof of Theorem 1**

The observation model for cell division process in each sample is given by

$$2\begin{bmatrix} \tilde{v}_{1,i}^{(0)} & \tilde{v}_{2,i}^{(0)} & \cdots & \tilde{v}_{M,i}^{(0)} \\ \tilde{v}_{1,i}^{(1)} & \tilde{v}_{2,i}^{(1)} & \cdots & \tilde{v}_{M,i}^{(1)} \\ \vdots & \vdots & \ddots & \vdots \\ \tilde{v}_{1,i}^{(NMS-2)} & \tilde{v}_{2,i}^{(NMS-2)} & \cdots & \tilde{v}_{M,i}^{(NMS-2)} \end{bmatrix} \begin{bmatrix} p_{1l} \\ p_{2l} \\ \vdots \\ p_{Ml} \end{bmatrix} + \begin{bmatrix} \eta_{l,i}^{(1)} \\ \eta_{l,i}^{(2)} \\ \vdots \\ \eta_{l,i}^{(NMS-1)} \end{bmatrix} = \begin{bmatrix} v_{l,i}^{(1)} \\ v_{l,i}^{(2)} \\ \vdots \\ v_{l,i}^{(NMS-1)} \end{bmatrix} \tag{S1}$$

where, $i \in \{1, N\}$. The equation (S1) is satisfied for each sample, $i$, and cell type, $l$. The goal is to find $p_{hl}$, $h, l \in [1, M]$, which fit the equations best in the minimum mean squared sense. We have

$$\hat{\mathbf{P}} = \arg\min_{\mathbf{P}} S(\mathbf{P}) \tag{S2-a}$$

s.t.

for $\forall\ 1 \leq h \leq M, 1 \leq l \leq M$,

$$0 \leq p_{hl} \leq 1, \tag{S2-b}$$

$$\sum_{l=1}^{M} p_{hl} = 1, \tag{S2-c}$$

where the objective function $S$ is defined as

$$S(\mathbf{P}) = \sum_{l=1}^{M} \sum_{i=1}^{N} \sum_{j=1}^{NMS-1} \left( v_{l,i}^{(j)} - 2\sum_{h=1}^{M} \tilde{v}_{h,i}^{(j-1)} p_{hl} \right)^2, \tag{S3}$$

The optimization problem in (S2) is convex, (cost function is quadratic and constraints are linear [2]). Using KKT (Karush–Kuhn–Tucker), the unconstrained solution for this problem is given by

$$S_U(\mathbf{P}) = \sum_{l=1}^{M} \left( S(\mathbf{P}_l) + \lambda_l \left( \sum_{h=1}^{M} p_{hl} - 1 \right) + \sum_{h=1}^{M} (\lambda'_{hl} - \lambda''_{hl}) p_{hl}, \right) \tag{S4}$$

where $\lambda'_{jl}$, $\lambda''_{jl}$ and $\lambda_l$ are Lagrange multipliers and $\mathbf{P}_l = [p_{1l} \ \cdots \ p_{Ml}]$. The value of $S(\mathbf{P}_l)$ can be simplified as

$$S(\mathbf{P}_l) = \sum_{i=1}^{N}\sum_{j=1}^{NMS-1}\left(v_{l,i}^{(j)}\right)^2 + 4\sum_{i=1}^{N}\sum_{j=1}^{NMS-1}\left(\sum_{h=1}^{M}\tilde{v}_{h,i}^{(j-1)}p_{hl}\right)^2 - 4\sum_{i=1}^{N}\sum_{j=1}^{NMS-1}v_{l,i}^{(j)}\sum_{h=1}^{M}\tilde{v}_{h,i}^{(j-1)}p_{hl}$$

$$= \sum_{i=1}^{N}\sum_{j=1}^{NMS-1}\left(v_{l,i}^{(j)}\right)^2 + 4\sum_{i=1}^{N}\sum_{j=1}^{NMS-1}\sum_{h=1}^{M}\sum_{k=1}^{M}\tilde{v}_{h,i}^{(j-1)}\tilde{v}_{k,i}^{(j-1)}p_{hl}p_{kl} - 4\sum_{i=1}^{N}\sum_{j=1}^{NMS-1}v_{l,i}^{(j)}\sum_{h=1}^{M}\tilde{v}_{h,i}^{(j-1)}p_{hl}$$

$$= \sum_{i=1}^{N}\sum_{j=1}^{NMS-1}\left(v_{l,i}^{(j)}\right)^2 + 4\sum_{j=1}^{NMS-1}\sum_{h=1}^{M}\left[\sum_{k=1}^{M}\sum_{i=1}^{N}\left(v_{h,i}^{(j-1)}v_{k,i}^{(j-1)}p_{hl}p_{kl} - \eta_{h,i}^{(j-1)}v_{k,i}^{(j-1)}p_{hl}p_{kl} - \eta_{k,i}^{(j-1)}v_{h,i}^{(j-1)}p_{hl}p_{kl} + \eta_{h,i}^{(j-1)}\eta_{k,i}^{(j-1)}p_{hl}p_{kl}\right)\right]$$

$$- 4\sum_{j=1}^{NMS-1}\sum_{i=1}^{N}v_{l,i}^{(j)}\sum_{h=1}^{M}v_{h,i}^{(j-1)}p_{hl} - \eta_{h,i}^{(j)}p_{hl},$$

$$= \sum_{i=1}^{N}\sum_{j=1}^{NMS-1}\left(v_{l,i}^{(j)}\right)^2 + 4\sum_{j=1}^{NMS-1}\sum_{h=1}^{M}\sum_{k=1}^{M}\sum_{i=1}^{N}v_{h,i}^{(j-1)}v_{k,i}^{(j-1)}p_{hl}p_{kl} - 4\sum_{j=1}^{NMS-1}\sum_{i=1}^{N}v_{l,i}^{(j)}\sum_{h=1}^{M}v_{h,i}^{(j-1)}p_{hl}$$

$$- 4\sum_{j=1}^{NMS-1}\sum_{h=1}^{M}\sum_{k=1}^{M}\sum_{i=1}^{N}\left(\eta_{h,i}^{(j-1)}v_{k,i}^{(j-1)}p_{hl}p_{kl} + \eta_{k,i}^{(j-1)}v_{h,i}^{(j-1)}p_{hl}p_{kl} - \eta_{h,i}^{(j-1)}\eta_{k,i}^{(j-1)}p_{hl}p_{kl}\right) + 2\sum_{j=1}^{NMS-1}\sum_{h=1}^{M}\sum_{i=1}^{N}v_{l,i}^{(j)}\eta_{h,i}^{(j)}p_{hl}$$

$$= \sum_{i=1}^{N}\sum_{j=1}^{NMS-1}\left(v_{l,i}^{(j)}\right)^2 + 4\sum_{i=1}^{N}\sum_{j=1}^{NMS-1}\left(\sum_{h=1}^{M}v_{h,i}^{(j-1)}p_{hl}\right)^2 - 4\sum_{j=1}^{NMS-1}\sum_{i=1}^{N}v_{l,i}^{(j)}\sum_{h=1}^{M}v_{h,i}^{(j-1)}p_{hl} + 4\sum_{j=1}^{NMS-1}\sum_{h=1}^{M}\sum_{k=1}^{M}\sum_{i=1}^{N}\eta_{h,i}^{(j-1)}\eta_{k,i}^{(j-1)}p_{hl}p_{kl}$$

$$- 4\sum_{j=1}^{NMS-1}\sum_{h=1}^{M}\sum_{i=1}^{N}\left(\eta_{h,i}^{(j-1)}\sum_{k=1}^{M}v_{k,i}^{(j-1)}p_{hl}p_{kl}\right) - 4\sum_{j=1}^{NMS-1}\sum_{k=1}^{M}\sum_{i=1}^{N}\left(\eta_{k,i}^{(j-1)}\sum_{h=1}^{M}v_{h,i}^{(j-1)}p_{hl}p_{kl}\right) + 4\sum_{j=1}^{NMS-1}\sum_{h=1}^{M}\sum_{i=1}^{N}v_{l,i}^{(j)}\eta_{h,i}^{(j)}p_{hl},$$

$$= \sum_{i=1}^{N}\sum_{j=1}^{NMS-1}\left(v_{l,i}^{(j)} - 2\sum_{h=1}^{M}v_{h,i}^{(j-1)}p_{hl}\right)^2 + 4\sum_{j=1}^{NMS-1}\sum_{h=1}^{M}\sum_{k=1}^{M}p_{hl}p_{kl}\sum_{i=1}^{N}\eta_{h,i}^{(j-1)}\eta_{k,i}^{(j-1)} - 4\sum_{j=1}^{NMS-1}\sum_{h=1}^{M}\sum_{k=1}^{M}p_{hl}p_{kl}\sum_{i=1}^{N}\eta_{h,i}^{(j-1)}v_{k,i}^{(j-1)}$$

$$+ 4\sum_{j=1}^{NMS-1}\sum_{h=1}^{M}p_{hl}\sum_{i=1}^{N}v_{l,i}^{(j)}\eta_{h,i}^{(j)}, \qquad (S5)$$

If either of the constraints in (S2-b) is inactive (is not satisfied with equality), the corresponding Lagrange multiplier is set to zero. And if either of them is active the value of $p_{jl}$ is obtained directly from (S2-b). Hence, here we solve the problem when the set of constraints in (S2-b) is not active, and compute the derivative of (S5) with respect to $p_{ql}$ and $\lambda_l$, $q,l \in [1,M]$. We have

$$\frac{\partial S_U(\mathbf{P})}{\partial p_{ql}} = -4\sum_{i=1}^{N}\sum_{j=1}^{NMS-1}v_{q,i}^{(j-1)}\left(v_{l,i}^{(j)} - 2\sum_{h=1}^{M}v_{h,i}^{(j-1)}p_{hl}\right) + 8\sum_{j=1}^{NMS-1}\sum_{h=1}^{M}p_{hl}\sum_{i=1}^{N}\eta_{q,i}^{(j-1)}\eta_{h,i}^{(j-1)}$$

$$- 8\sum_{j=1}^{NMS-1}\sum_{h=1}^{M}p_{hl}\sum_{i=1}^{N}\eta_{q,i}^{(j-1)}v_{h,i}^{(j-1)} + 4\sum_{j=1}^{NMS-1}\sum_{i=1}^{N}v_{q,i}^{(j)}\eta_{l,i}^{(j)} + \lambda_l, \qquad (S6\text{-a})$$

$$\frac{\partial S_U(\mathbf{P})}{\partial \lambda_l} = \left(\sum_{j=1}^{M}p_{jl} - 1\right), \qquad (S6\text{-b})$$

By dividing both sides of (S6-a) by $N$ and because the noise of different samples are centered processes, we have

$$\frac{1}{N}\frac{\partial S(\mathbf{P}_l)}{\partial p_{ql}} = -\frac{4}{N}\left(\sum_{i=1}^{N}\sum_{j=1}^{NMS-1} v_{q,i}^{(j-1)}\left(v_{l,i}^{(j)} - 2\sum_{h=1}^{M} v_{h,i}^{(j-1)} p_{hl}\right) - 2\sum_{j=1}^{NMS-1}\sum_{h=1}^{M} p_{hl}\sum_{i=1}^{N}\eta_{q,i}^{(j-1)}\eta_{h,i}^{(j-1)} + \frac{\lambda_l}{4}\right)$$
$$= -\frac{4}{N}\left(0.5\sum_{i=1}^{N}\sum_{j=1}^{NMS-1} v_{q,i}^{(j-1)} v_{l,i}^{(j)} - \sum_{h=1}^{M}\left(v_{q,i}^{(j-1)} v_{h,i}^{(j-1)} + \eta_{q,i}^{(j-1)}\eta_{h,i}^{(j-1)}\right) p_{hl} + \frac{\lambda_l}{4}\right)$$
(S7)

By writing (S7) in matrix form, we have

$$\left[\sum_{i=1}^{N}\sum_{j=1}^{NMS-1}\left(v_{q,i}^{(j-1)} v_{1,i}^{(j-1)} + \eta_{q,i}^{(j-1)}\eta_{1,i}^{(j-1)}\right) \cdots \sum_{i=1}^{N}\sum_{j=1}^{NMS-1}\left(v_{q,i}^{(j-1)} v_{M,i}^{(j-1)} + \eta_{q,i}^{(j-1)}\eta_{M,i}^{(j-1)}\right)\right]\begin{bmatrix}p_{1l}\\ \vdots \\ p_{Ml}\end{bmatrix} = 0.5\sum_{i=1}^{N}\sum_{j=1}^{NMS-1} v_{q,i}^{(j-1)} v_{l,i}^{(j)} - \frac{\lambda_l}{4}$$
(S8)

If we write the above equation for all values of $q$, we have following equations

$$\begin{bmatrix}\sum_{j=1}^{NMS-1}\sum_{i=1}^{N}\left(v_{1,i}^{(j-1)} v_{1,i}^{(j-1)} + \eta_{1,i}^{(j-1)}\eta_{1,i}^{(j-1)}\right) & \cdots & \sum_{j=1}^{NMS-1}\sum_{i=1}^{N}\left(v_{1,i}^{(j-1)} v_{M,i}^{(j-1)} + \eta_{1,i}^{(j-1)}\eta_{M,i}^{(j-1)}\right) \\ \vdots & \ddots & \vdots \\ \sum_{j=1}^{NMS-1}\sum_{i=1}^{N}\left(v_{M,i}^{(j-1)} v_{1,i}^{(j-1)} + \eta_{M,i}^{(j-1)}\eta_{1,i}^{(j-1)}\right) & \cdots & \sum_{j=1}^{NMS-1}\sum_{i=1}^{N}\left(v_{M,i}^{(j-1)} v_{M,i}^{(j-1)} + \eta_{M,i}^{(j-1)}\eta_{M,i}^{(j-1)}\right)\end{bmatrix}\begin{bmatrix}p_{1l}\\ \vdots \\ p_{Ml}\end{bmatrix} = 0.5\begin{bmatrix}\sum_{i=1}^{N}\sum_{j=1}^{NMS-1} v_{1,i}^{(j-1)} v_{l,i}^{(j)} \\ \vdots \\ \sum_{i=1}^{N}\sum_{j=1}^{NMS-1} v_{M,i}^{(j-1)} v_{l,i}^{(j)}\end{bmatrix} - \frac{\lambda_l}{4}$$
(S9)

Hence, $\mathbf{P}_l$ is given by

$$\begin{bmatrix}p_{1l}\\ \vdots \\ p_{Ml}\end{bmatrix} = 0.5\left(\begin{bmatrix}\sum_{j=1}^{NMS-1}\sum_{i=1}^{N} v_{1,i}^{(j-1)} v_{1,i}^{(j-1)} & \cdots & \sum_{j=1}^{NMS-1}\sum_{i=1}^{N} v_{1,i}^{(j-1)} v_{M,i}^{(j-1)} \\ \vdots & \ddots & \vdots \\ \sum_{j=1}^{NMS-1}\sum_{i=1}^{N} v_{M,i}^{(j-1)} v_{1,i}^{(j-1)} & \cdots & \sum_{j=1}^{NMS-1}\sum_{i=1}^{N} v_{M,i}^{(j-1)} v_{M,i}^{(j-1)}\end{bmatrix} + \begin{bmatrix}\sum_{j=1}^{NMS-1}\sum_{i=1}^{N}\eta_{1,i}^{(j-1)}\eta_{1,i}^{(j-1)} & \cdots & \sum_{j=1}^{NMS-1}\sum_{i=1}^{N}\eta_{1,i}^{(j-1)}\eta_{M,i}^{(j-1)} \\ \vdots & \ddots & \vdots \\ \sum_{j=1}^{NMS-1}\sum_{i=1}^{N}\eta_{M,i}^{(j-1)}\eta_{1,i}^{(j-1)} & \cdots & \sum_{j=1}^{NMS-1}\sum_{i=1}^{N}\eta_{M,i}^{(j-1)}\eta_{M,i}^{(j-1)}\end{bmatrix}\right)^{-1}$$
$$\begin{bmatrix}\sum_{i=1}^{N}\sum_{j=1}^{MMS-1} v_{1,i}^{(j-1)} v_{l,i}^{(j)} - \frac{\lambda_l}{4} \\ \vdots \\ \sum_{i=1}^{N}\sum_{j=1}^{NMS-1} v_{M,i}^{(j-1)} v_{l,i}^{(j)} - \frac{\lambda_l}{4}\end{bmatrix}$$
(S10)

If the variance of noise for each type of cell in all samples is considered equal, i.e., $\sigma_{ji}^2 = \sigma_j^2$, we have

$$\begin{bmatrix} p_{1l} \\ \vdots \\ p_{Ml} \end{bmatrix} = 0.5 \left( \begin{bmatrix} \sum_{j=1}^{NMS-1}\sum_{i=1}^{N} v_{1,i}^{(j-1)}v_{1,i}^{(j-1)} & \cdots & \sum_{j=1}^{NMS-1}\sum_{i=1}^{N} v_{1,i}^{(j-1)}v_{M,i}^{(j-1)} \\ \vdots & \ddots & \vdots \\ \sum_{j=1}^{NMS-1}\sum_{i=1}^{N} v_{M,i}^{(j-1)}v_{1,i}^{(j-1)} & \cdots & \sum_{j=1}^{NMS-1}\sum_{i=1}^{N} v_{M,i}^{(j-1)}v_{M,i}^{(j-1)} \end{bmatrix} + N(NMS-1)\begin{bmatrix} \sigma_1^2 & \cdots & 0 \\ \vdots & \ddots & \vdots \\ 0 & \cdots & \sigma_M^2 \end{bmatrix} \right)^{-1}$$

$$\begin{bmatrix} \sum_{i=1}^{N}\sum_{j=1}^{NMS-1} v_{1,i}^{(j-1)}v_{l,i}^{(j)} - \frac{\lambda_l}{4} \\ \vdots \\ \sum_{i=1}^{N}\sum_{j=1}^{NMS-1} v_{M,i}^{(j-1)}v_{l,i}^{(j)} - \frac{\lambda_l}{4} \end{bmatrix}$$
(S11)

which can be simplified to

$$\hat{\mathbf{P}}_l = 0.5\left( \left(\mathbf{V}_{:,1:N}^{(0:NMS-2)}\right)^{\dagger}\left(\mathbf{V}_{:,1:N}^{(0:NMS-2)}\right) + N(NMS-1)\mathbf{R}_n \right)^{-1}\left(\mathbf{V}_{:,1:N}^{(0:NMS-1)}\right)^{\dagger}\mathbf{V}_{l,(1:N)}^{(1:NMS-1)} - \Lambda_l,$$
(S12)

where,

$$\mathbf{V}_{:,1:N}^{(0:NMS-2)} = \begin{bmatrix} v_{1,1}^{(0)} & \cdots & v_{M,1}^{(0)} \\ \vdots & \ddots & \vdots \\ v_{1,1}^{(NMS-2)} & \cdots & v_{M,1}^{(NMS-2)} \\ \vdots & \vdots & \vdots \\ v_{1,N}^{(0)} & \cdots & v_{M,N}^{(0)} \\ \vdots & \ddots & \vdots \\ v_{1,N}^{(NMS-2)} & \cdots & v_{M,N}^{(NMS-2)} \end{bmatrix},$$
(S13)

and

$$\mathbf{V}_{l,(1:N)}^{(1:NMS-1)} = \begin{bmatrix} v_{l,1}^{(1)} & \cdots & v_{l,1}^{(M)} & \cdots & v_{l,N}^{(1)} & \cdots & v_{l,N}^{(NMS-1)} \end{bmatrix}^{\dagger}$$
(S14)

and

$$\mathbf{R}_n = \begin{bmatrix} \sigma_1^2 & \cdots & 0 \\ \vdots & \ddots & \vdots \\ 0 & \cdots & \sigma_M^2 \end{bmatrix}$$
(S15)

and

$$\Lambda_l = \begin{bmatrix} \frac{\lambda_l}{4} & \cdots & \frac{\lambda_l}{4} \end{bmatrix}^{\dagger}$$
(S16)

where $\lambda_l$ is obtained from the following equation for each value of $l$

$$\sum_{h=1}^{M} p_{hl} = 1 \tag{S17}$$

As stated, in the above derivations we did not explicitly incorporated the constraints in (S2-b) and assumed they are inactive. If the assumption is not valid, we will arrive at values of $p_{hl}$ possibly greater than one or negative. In this case, we enforce the violated constraint with equality and solve the optimization problem again. In case, there are multiple such violated constraints, this process is repeated for different combinations of enforced constraints (see [3] pp.314-357 for details).

**Section C**

In this Section, we derive the ML estimator for the transition probability of the cell proliferation Markov chain with observations made in the presence of additive white Gaussian noise. In the first step, using a joint probability formula for $\hat{\mathbf{P}}_{ML}$, we have

$$\begin{aligned}
\hat{\mathbf{P}}_{ML} &= \max_{\mathbf{P}} P\left(v_{1,1}^{(0)},...,v_{1,N}^{(0)},...,v_{M,1}^{(0)},...,v_{M,N}^{(0)},...,v_{1,1}^{(NMS-1)},...,v_{1,N}^{(NMS-1)},...,v_{M,N}^{(NMS-1)},...,v_{M,N}^{(NMS-1)}\Big|\mathbf{P}\right) \\
&\stackrel{(a)}{=} \max_{\mathbf{P}} P\left(v_{1,1}^{(0)},...,v_{1,N}^{(0)},...,v_{M,1}^{(0)},...,v_{M,N}^{(0)}\Big|\mathbf{P}\right) \\
&\quad P\left(v_{1,1}^{(1)},...,v_{1,N}^{(1)},...,v_{M,1}^{(1)},...,v_{M,N}^{(1)}\Big|\mathbf{P},v_{1,1}^{(0)},...,v_{1,N}^{(0)},...,v_{M,1}^{(0)},...,v_{M,N}^{(0)}\right)... \\
&\quad P\left(v_{1,1}^{(NMS-1)},...,v_{1,N}^{(NMS-1)},...,v_{M,1}^{(NMS-1)},...,v_{M,N}^{(NMS-1)}\Big|\mathbf{P},v_{1,1}^{(0)},...,v_{1,N}^{(0)},...,v_{M,1}^{(0)},...,v_{M,N}^{(0)},...,\right.\\
&\quad \left. v_{1,1}^{(NMS-2)},...,v_{1,N}^{(NMS-2)},...,v_{M,1}^{(NMS-2)},...,v_{M,N}^{(NMS-2)}\right) \\
&\stackrel{(b)}{=} \max_{\mathbf{P}} P\left(v_{1,1}^{(0)},...,v_{1,N}^{(0)},...,v_{M,1}^{(0)},...,v_{M,N}^{(0)}\Big|\mathbf{P}\right) \\
&\quad P\left(v_{1,1}^{(1)},...,v_{1,N}^{(1)},...,v_{M,1}^{(1)},...,v_{M,N}^{(1)}\Big|\mathbf{P},v_{1,1}^{(0)},...,v_{1,N}^{(0)},...,v_{M,1}^{(0)},...,v_{M,N}^{(0)}\right)... \\
&\quad P\left(v_{1,1}^{(NMS-1)},...,v_{1,N}^{(NMS-1)},...,v_{M,1}^{(NMS-1)},...,v_{M,N}^{(NMS-1)}\Big|\mathbf{P},v_{1,1}^{(NMS-2)},...,v_{1,N}^{(NMS-2)},...,v_{M,1}^{(NMS-2)},...,v_{M,N}^{(NMS-2)}\right),
\end{aligned} \tag{S18}$$

(a) is obtained using the chain rule and (b) is due to first order Markov property of population size. Moreover, it is assumed that the observations are independent, hence, in (S18) we have

$$P\left(v_{1,1}^{(0)},...,v_{1,N}^{(0)},...,v_{M,1}^{(0)},...,v_{M,N}^{(0)}\Big|\mathbf{P}\right) = \prod_{i=1}^{N}\prod_{j=1}^{M}P\left(v_{j,i}^{(0)}\Big|\mathbf{P}\right)$$

$$= \prod_{i=1}^{N}\prod_{j=1}^{M}\frac{1}{\sqrt{2\pi}\sigma_{ji}}\exp\left(\frac{-\left(v_{j,i}^{(0)}-\tilde{v}_{j,i}^{(0)}\right)^2}{2\sigma_{ji}^2}\right) \quad (S19)$$

$$= \frac{1}{(2\pi)^{MN/2}\prod_{i=1}^{N}\prod_{j=1}^{M}\sigma_{ji}}\exp\left(-\sum_{i=1}^{N}\sum_{j=1}^{M}\frac{\left(v_{j,i}^{(0)}-\tilde{v}_{j,i}^{(0)}\right)^2}{2\sigma_{ji}^2}\right),$$

$$P\left(v_{1,1}^{(1)},...,v_{1,N}^{(1)},...,v_{M,1}^{(1)},...,v_{M,N}^{(1)}\Big|\mathbf{P},v_{1,1}^{(0)},...,v_{1,N}^{(0)},...,v_{M,1}^{(0)},...,v_{M,N}^{(0)}\right) = \prod_{i=1}^{N}\prod_{j=1}^{M}P\left(v_{j,i}^{(1)}\Big|\mathbf{P},v_{1,1}^{(0)},...,v_{1,N}^{(0)},...,v_{M,1}^{(0)},...,v_{M,N}^{(0)}\right)$$

$$= \prod_{i=1}^{N}\prod_{j=1}^{M}\frac{1}{\sqrt{2\pi}\sigma_{ji}}\exp\left(-\frac{\left(v_{j,i}^{(1)}-2\sum_{h=1}^{M}\tilde{v}_{h,i}^{(0)}p_{hj}\right)^2}{2\sigma_{ji}^2}\right)$$

$$= \frac{1}{(2\pi)^{MN/2}\prod_{i=1}^{N}\prod_{j=1}^{M}\sigma_{ji}}\exp\left(-\sum_{i=1}^{N}\sum_{j=1}^{M}\frac{\left(v_{j,i}^{(1)}-2\sum_{h=1}^{M}\tilde{v}_{h,i}^{(0)}p_{hj}\right)^2}{2\sigma_{ji}^2}\right) \quad (S20)$$

$$= \frac{1}{(2\pi)^{MN/2}\prod_{i=1}^{N}\prod_{j=1}^{M}\sigma_{ji}}\exp\left(-\sum_{i=1}^{N}\sum_{j=1}^{M}\frac{\left(v_{j,i}^{(1)}-2\sum_{h=1}^{M}\tilde{v}_{h,i}^{(0)}p_{hj}\right)^2}{2\sigma_{ji}^2}\right)$$

$$P\left(v_{1,1}^{(NMS-1)},...,v_{1,N}^{(NMS-1)},...,v_{M,1}^{(1)},...,v_{M,N}^{(1)}\Big|\mathbf{P},v_{1,1}^{(NMS-2)},...,v_{1,N}^{(NMS-2)},...,v_{M,1}^{(NMS-2)},...,v_{M,N}^{(NMS-2)}\right) =$$

$$\prod_{i=1}^{N}\prod_{j=1}^{M}P\left(v_{j,i}^{(NMS-1)}\Big|\mathbf{P},v_{1,1}^{(NMS-2)},...,v_{1,N}^{(NMS-2)},...,v_{M,1}^{(NMS-2)},...,v_{M,N}^{(NMS-2)}\right) =$$

$$\prod_{i=1}^{N}\prod_{j=1}^{M}\frac{1}{\sqrt{2\pi}\sigma_{ji}}\exp\left(\frac{-\left(v_{j,i}^{(NMS-1)}-2\sum_{h=1}^{M}\tilde{v}_{h,i}^{(NMS-2)}p_{hj}\right)^2}{2\sigma_{ji}^2}\right) =$$

$$\frac{1}{(2\pi)^{MN/2}\prod_{i=1}^{N}\prod_{j=1}^{M}\sigma_{ji}}\exp\left(-\sum_{i=1}^{N}\sum_{j=1}^{M}\frac{\left(v_{j,i}^{(NMS-1)}-2\sum_{h=1}^{M}\tilde{v}_{h,i}^{(NMS-2)}p_{hj}\right)^2}{2\sigma_{ji}^2}\right). \quad (S21)$$

where, $\sigma_{ji}^2$, $j \in [1, M]$ denotes the standard deviation of Gaussian noise for cell type $j$ in sample $i$. Replacing (S19)-(S21) in (S18), we have

$$P\left(v_{1,1}^{(0)}, ..., v_{1,N}^{(0)}, ..., v_{M,1}^{(0)}, ..., v_{M,N}^{(0)}, ..., v_{1,1}^{(NMS-1)}, ..., v_{1,N}^{(NMS-1)}, ..., v_{M,N}^{(NMS-1)}, ..., v_{M,N}^{(NMS-1)} \middle| \mathbf{P}\right) =$$

$$\frac{1}{(2\pi)^{(MN)NMS/2} \left(\prod_{i=1}^{N}\prod_{j=1}^{M} \sigma_{ji}\right)^{NMS}}$$

$$\exp\left(-\sum_{i=1}^{N}\sum_{j=1}^{M} \frac{\left(v_{j,i}^{(0)} - \tilde{v}_{j,i}^{(0)}\right)^2}{2\sigma_{ji}^2} - \sum_{i=1}^{N}\sum_{j=1}^{M} \frac{\left(v_{j,i}^{(1)} - 2\sum_{h=1}^{M}\tilde{v}_{h,i}^{(0)} p_{hj}\right)^2}{2\sigma_{ji}^2} - ... - \sum_{i=1}^{N}\sum_{j=1}^{M} \frac{\left(v_{j,i}^{(NMS-1)} - 2\sum_{h=1}^{M}\tilde{v}_{h,i}^{(NMS-2)} p_{hj}\right)^2}{2\sigma_{ji}^2}\right) \quad (S22)$$

Taking the Logarithm of the above equation and ignoring the constant terms, we have

$$\log P\left(v_{1,1}^{(0)}, ..., v_{1,N}^{(0)}, ..., v_{M,1}^{(0)}, ..., v_{M,N}^{(0)}, ..., v_{1,1}^{(NMS-1)}, ..., v_{1,N}^{(NMS-1)}, ..., v_{M,N}^{(NMS-1)}, ..., v_{M,N}^{(NMS-1)} \middle| \mathbf{P}\right) =$$

$$-NMS \log\left(\left(\sqrt{2\pi}\right)^{MN} \prod_{i=1}^{N}\prod_{j=1}^{M} \sigma_{ji}\right) - \sum_{i=1}^{N}\sum_{j=1}^{M} \frac{\left(v_{j,i}^{(0)} - \tilde{v}_{j,i}^{(0)}\right)^2}{2\sigma_{ji}^2} - \sum_{i=1}^{N}\sum_{j=1}^{M} \frac{\left(v_{j,i}^{(1)} - 2\sum_{i=1}^{M}\tilde{v}_{h,i}^{(0)} p_{hj}\right)^2}{2\sigma_{ji}^2} - ... \quad (S23)$$

$$-\sum_{i=1}^{N}\sum_{j=1}^{M} \frac{\left(v_{j,i}^{(NMS-1)} - 2\sum_{i=1}^{M}\tilde{v}_{h,i}^{(NMS-2)} p_{hj}\right)^2}{2\sigma_{ji}^2}.$$

Taking the derivative with respect to $p_{kl}$ and setting it to zero, we have $M$ equations as follows

$$\sum_{i=1}^{N}\left(\frac{\tilde{v}_{k,i}^{(0)}\left(v_{l,i}^{(1)} - 2\sum_{h=1}^{M}\tilde{v}_{h,i}^{(0)} p_{hl}\right)}{2\sigma_{li}^2} + ... + \frac{\tilde{v}_{k,i}^{(NMS-2)}\left(v_{l,i}^{(NMS-1)} - 2\sum_{i=1}^{M}\tilde{v}_{h,i}^{(NMS-2)} p_{hl}\right)}{2\sigma_{li}^2}\right) = 0, \Rightarrow$$

$$2\sum_{i=1}^{N}\sum_{j=1}^{NMS-1}\sum_{h=1}^{M} \frac{\tilde{v}_{k,i}^{(j-1)}\tilde{v}_{h,i}^{(j-1)} p_{hl}}{2\sigma_{li}^2} = \sum_{i=1}^{N}\sum_{j=1}^{NMS-1} \frac{\tilde{v}_{k,i}^{(j-1)} v_{l}^{(j)}}{2\sigma_{li}^2}, \Rightarrow$$

$$\sum_{i=1}^{N}\begin{bmatrix}\tilde{v}_{k,i}^{(0)} & \tilde{v}_{k,i}^{(1)} & \cdots & \tilde{v}_{k,i}^{(NMS-2)}\end{bmatrix}\begin{bmatrix} 2\sigma_{li}^{-2}\sum_{h=1}^{M}\tilde{v}_{h,i}^{(0)} p_{hl} \\ 2\sigma_{li}^{-2}\sum_{h=1}^{M}\tilde{v}_{h,i}^{(1)} p_{hl} \\ \vdots \\ 2\sigma_{li}^{-2}\sum_{h=1}^{M}\tilde{v}_{h,i}^{(NMS-2)} p_{hl}\end{bmatrix} = \sum_{i=1}^{N}\begin{bmatrix}\tilde{v}_{k,i}^{(0)} & \tilde{v}_{k,i}^{(1)} & \cdots & \tilde{v}_{k,i}^{(NMS-2)}\end{bmatrix}\begin{bmatrix} \sigma_{li}^{-2} v_{l,i}^{(1)} \\ \sigma_{li}^{-2} v_{l,i}^{(2)} \\ \vdots \\ \sigma_{li}^{-2} v_{l,i}^{(NMS-1)}\end{bmatrix}. \quad (S24)$$

In the setting under consideration, we make separate observations based on each sample, and attempt to obtain a common optimized transition probability for the underlying Markov process. As a result, we satisfy the above equation by enforcing the constraint for each sample. We have

$$\begin{bmatrix} \tilde{v}_{k,i}^{(0)} & \tilde{v}_{k,i}^{(1)} & \cdots & \tilde{v}_{k,i}^{(NMS-2)} \end{bmatrix} \begin{bmatrix} 2\sigma_{li}^{-2} \sum_{h=1}^{M} \tilde{v}_{h,i}^{(0)} p_{hl} \\ 2\sigma_{li}^{-2} \sum_{h=1}^{M} \tilde{v}_{h,i}^{(1)} p_{hl} \\ \vdots \\ 2\sigma_{li}^{-2} \sum_{h=1}^{M} \tilde{v}_{h,i}^{(NMS-2)} p_{hl} \end{bmatrix} = \begin{bmatrix} \tilde{v}_{k,i}^{(0)} & \tilde{v}_{k,i}^{(1)} & \cdots & \tilde{v}_{k,i}^{(NMS-2)} \end{bmatrix} \begin{bmatrix} \sigma_{li}^{-2} v_{l,i}^{(1)} \\ \sigma_{li}^{-2} v_{l,i}^{(2)} \\ \vdots \\ \sigma_{li}^{-2} v_{l,i}^{(NMS-1)} \end{bmatrix}. \tag{S25}$$

Hence for all values of $k$, we have following equations,

$$\begin{bmatrix} \tilde{v}_{1,i}^{(0)} & \tilde{v}_{1,i}^{(1)} & \cdots & \tilde{v}_{1,i}^{(NMS-2)} \\ \tilde{v}_{2,i}^{(0)} & \tilde{v}_{2,i}^{(1)} & \cdots & \tilde{v}_{2,i}^{(NMS-2)} \\ \vdots & \vdots & \ddots & \vdots \\ \tilde{v}_{M,i}^{(0)} & \tilde{v}_{M,i}^{(1)} & \cdots & \tilde{v}_{M,i}^{(NMS-2)} \end{bmatrix} \begin{bmatrix} 2\sigma_{li}^{-2} \sum_{h=1}^{M} \tilde{v}_{h,i}^{(0)} p_{il} \\ 2\sigma_{li}^{-2} \sum_{h=1}^{M} \tilde{v}_{h,i}^{(1)} p_{il} \\ \vdots \\ 2\sigma_{li}^{-2} \sum_{h=1}^{M} \tilde{v}_{h,i}^{(NMS-2)} p_{il} \end{bmatrix} = \begin{bmatrix} \tilde{v}_{1,i}^{(0)} & \tilde{v}_{1,i}^{(1)} & \cdots & \tilde{v}_{1,i}^{(NMS-2)} \\ \tilde{v}_{2,i}^{(0)} & \tilde{v}_{2,i}^{(1)} & \cdots & \tilde{v}_{2,i}^{(NMS-2)} \\ \vdots & \vdots & \ddots & \vdots \\ \tilde{v}_{M,i}^{(0)} & \tilde{v}_{M,i}^{(1)} & \cdots & \tilde{v}_{M,i}^{(NMS-2)} \end{bmatrix} \begin{bmatrix} \sigma_{li}^{-2} v_{l,i}^{(1)} \\ \sigma_{li}^{-2} v_{l,i}^{(2)} \\ \vdots \\ \sigma_{li}^{-2} v_{l,i}^{(NMS-1)} \end{bmatrix}, \tag{S26}$$

If matrix $\mathbf{V}_i$ is full column rank, we simply have

$$\begin{bmatrix} 2\sum_{h=1}^{M} \tilde{v}_{h,i}^{(0)} p_{il} \\ 2\sum_{h=1}^{M} \tilde{v}_{h,i}^{(1)} p_{il} \\ \vdots \\ 2\sum_{h=1}^{M} \tilde{v}_{h,i}^{(NMS-2)} p_{il} \end{bmatrix} = \begin{bmatrix} v_{l,i}^{(1)} \\ v_{l,i}^{(2)} \\ \vdots \\ v_{l,i}^{(NMS-1)} \end{bmatrix} \Rightarrow$$

$$2 \begin{bmatrix} \tilde{v}_{1,i}^{(0)} & \tilde{v}_{2,i}^{(0)} & \cdots & \tilde{v}_{M,i}^{(0)} \\ v_{1,i}^{(1)} & v_{2,i}^{(1)} & \cdots & \tilde{v}_{M,i}^{(1)} \\ \vdots & \vdots & \ddots & \vdots \\ v_{1,i}^{(NMS-2)} & v_{2,i}^{(NMS-2)} & \cdots & \tilde{v}_{M,i}^{(NMS-2)} \end{bmatrix} \begin{bmatrix} p_{1l} \\ p_{2l} \\ \vdots \\ p_{Ml} \end{bmatrix} = \begin{bmatrix} v_{l,i}^{(1)} \\ v_{l,i}^{(2)} \\ \vdots \\ v_{l,i}^{(NMS-1)} \end{bmatrix}. \tag{S27}$$

For each value of $i \in [1, N]$, if $NMS - 1 < M$, the above system of linear equations for $p_{1l}$ has infinitely many solutions, and is an underdetermined system. If $NMS - 1 = M$, the above system of linear equations has a single unique solution. Due to independency of noise in different samples, joint probability of observations in (18), can be factorized to multiplication of joint probability for $N$ samples. Hence, the solution of (S27) maximizes each component of joint probability in (S18). If $NMS - 1 > M$, such a system has no solution, and is an over determined system. In this case, if we solve (S27) with $\ell_2$ norm, we obtain the same results of Theorem 1.

**Section D**

In a Poisson process, the number of observed occurrences fluctuates about its mean $\tilde{v}_{j,i}^{(k)}$ with a standard deviation of $\sigma_{j,i} = \tilde{v}_{j,i}^{(k)}$. These fluctuations are due to what is known as *shot noise* and are signal dependent. In this Section, the system of equations for cell proliferation when observations are corrupted by shot noise is derived using an ML estimator. In shot noise, the observations are Poisson distributed, and the terms in (S18) may be computed as follows

$$P\left(v_{1,1}^{(0)},...,v_{1,N}^{(0)},...,v_{M,1}^{(0)},...,v_{M,N}^{(0)}\Big|\mathbf{P}\right) = \prod_{i=1}^{N}\prod_{j=1}^{M} P\left(v_{j,i}^{(0)}\Big|\mathbf{P}\right)$$

$$= \prod_{i=1}^{N}\prod_{j=1}^{M} \frac{\left(\tilde{v}_{j,i}^{(0)}\right)^{v_{j}^{(0)}}}{v_{j,i}^{(0)}!} \exp(-\tilde{v}_{j,i}^{(0)}) \quad (S28)$$

$$P\left(v_{1,1}^{(1)},...,v_{1,N}^{(1)},...,v_{M,1}^{(1)},...,v_{M,N}^{(1)}\Big|\mathbf{P},v_{1,1}^{(0)},...,v_{1,N}^{(0)},...,v_{M,1}^{(0)},...,v_{M,N}^{(0)}\right)$$

$$= \prod_{i=1}^{N}\prod_{j=1}^{M} P\left(v_{j,i}^{(1)}\Big|\mathbf{P},v_{1,1}^{(0)},...,v_{1,N}^{(0)},...,v_{M,1}^{(0)},...,v_{M,N}^{(0)}\right)$$

$$= \prod_{i=1}^{N}\prod_{j=1}^{M} \frac{\left(2\sum_{h=1}^{M}\tilde{v}_{h,i}^{(0)} p_{hj}\right)^{v_{j,i}^{(1)}}}{v_{j,i}^{(1)}!} \exp(-2\sum_{h=1}^{M}\tilde{v}_{h,i}^{(0)} p_{hj}) \quad (S29)$$

$$P\left(v_{1,1}^{(NMS-1)},...,v_{1,N}^{(NMS-1)},...,v_{M,1}^{(NMS-1)},...,v_{M,N}^{(NMS-1)}\Big|\mathbf{P},v_{1,1}^{(NMS-2)},...,v_{1,N}^{(NMS-2)},...,v_{M,1}^{(NMS-2)},...,v_{M,N}^{(NMS-2)}\right)$$

$$= \prod_{i=1}^{N}\prod_{j=1}^{M} P\left(v_{j,i}^{(NMS-1)}\Big|\mathbf{P},v_{1,1}^{(NMS-2)},...,v_{1,N}^{(NMS-2)},...,v_{M,1}^{(NMS-2)},...,v_{M,N}^{(NMS-2)}\right)$$

$$= \prod_{i=1}^{N}\prod_{j=1}^{M} \frac{\left(2\sum_{h=1}^{M}\tilde{v}_{h,i}^{(NMS-2)} p_{ij}\right)^{v_{ji}^{(NMS-1)}}}{v_{j,i}^{(NMS-1)}!} \exp(-2\sum_{h=1}^{M}\tilde{v}_{h,i}^{(NMS-2)} p_{hj}) \quad (S30)$$

in which $\tilde{v}_{j,i}^{(0)}$ is the true initial population size of cell type $j$ in sample $i$. Hence, using (S28) – (S30) in (S18), we have

$$P\left(v_{1,1}^{(0)},...,v_{1,N}^{(0)},...,v_{M,N}^{(0)},...,v_{M,N}^{(NMS-1)},...,v_{M,N}^{(NMS-1)}\Big|\mathbf{P}\right) = \prod_{i=1}^{N}\prod_{j=1}^{M}\frac{\left(\tilde{v}_{j,i}^{(0)}\right)^{v_{j,i}^{(0)}}}{v_{j,i}^{(0)}!}\prod_{i=1}^{N}\prod_{j=1}^{M}\frac{\left(2\sum_{h=1}^{M}\tilde{v}_{h,i}^{(0)}p_{hj}\right)^{v_{j,i}^{(1)}}}{v_{j,i}^{(1)}!}...$$

$$\prod_{i=1}^{N}\prod_{j=1}^{M}\frac{\left(2\sum_{h=1}^{M}\tilde{v}_{h,i}^{(NMS-2)}p_{ij}\right)^{v_{ji}^{(M)}}}{v_{j,i}^{(NMS-1)}!} \quad \text{(S31)}$$

$$\prod_{i=1}^{N}\prod_{j=1}^{M}\exp\left(-\left(\tilde{v}_{j,i}^{(0)} + 2\sum_{i=1}^{M}\tilde{v}_{h,i}^{(0)}p_{hj} + ... + 2\sum_{i=1}^{M}\tilde{v}_{h,i}^{(NMS-2)}p_{hj}\right)\right)$$

Taking the Logarithm of the above equation and ignoring the constant terms, we have

$$\log P\left(v_{1,1}^{(0)},...,v_{1,N}^{(0)},...,v_{M,N}^{(0)},...,v_{M,N}^{(NMS-1)},...,v_{M,N}^{(NMS-1)}\Big|\mathbf{P}\right) =$$

$$\sum_{i=1}^{N}\sum_{j=1}^{M}\left(v_{j,i}^{(0)}\log_e \tilde{v}_{j,i}^{(0)} - \log_e v_{j,i}^{(0)}! + v_{j,i}^{(1)}\log_e 2\sum_{h=1}^{M}\tilde{v}_{h,i}^{(0)}p_{hj} - \log_e v_{j,i}^{(1)}!... + v_{j,i}^{(NMS-1)}\log_e 2\sum_{h=1}^{M}\tilde{v}_{h,i}^{(NMS-2)}p_{hj} - \log_e v_{j,i}^{(NMS-1)}!\right. \quad \text{(S32)}$$

$$\left. -\tilde{v}_{j,i}^{(0)} - 2\sum_{i=1}^{M}\tilde{v}_{h,i}^{(0)}p_{hj} - ... - 2\sum_{i=1}^{M}\tilde{v}_{h,i}^{(NMS-2)}p_{hj}\right)$$

Computing the derivative with respect to $p_{kl}$ and setting the result to zero, we have $M$ equations

$$\sum_{i=1}^{N}\frac{v_{l,i}^{(1)}\tilde{v}_{k,i}^{(0)}}{\sum_{h=1}^{M}\tilde{v}_{h,i}^{(0)}p_{hl}} + ... + \frac{v_{l,i}^{(NMS-1)}\tilde{v}_{k,i}^{(NMS-2)}}{\sum_{h=1}^{M}\tilde{v}_{h,i}^{(NMS-2)}p_{hl}} - 2\left(\tilde{v}_{k,i}^{(0)} + ... + \tilde{v}_{k,i}^{(NMS-2)}\right) = 0 \Rightarrow$$

$$\sum_{i=1}^{N}\frac{v_{l,i}^{(1)}\tilde{v}_{k,i}^{(0)}}{\sum_{h=1}^{M}\tilde{v}_{h,i}^{(0)}p_{hl}} + ... + \frac{v_{l,i}^{(NMS-1)}\tilde{v}_{k,i}^{(NMS-2)}}{\sum_{h=1}^{M}\tilde{v}_{h,i}^{(NMS-2)}p_{hl}} = \sum_{i=1}^{N}2\left(\tilde{v}_{k,i}^{(0)} + ... + \tilde{v}_{k,i}^{(NMS-2)}\right) \Rightarrow$$

$$\sum_{i=1}^{N}\left(v_{l,i}^{(1)}\tilde{v}_{k,i}^{(0)}\prod_{\substack{q=1\\(q,j)\neq(i,0)}}^{N}\prod_{j=0,}^{NMS-2}\sum_{h=1}^{M}\tilde{v}_{h,i}^{(j)}p_{hl} + ... + v_{l,i}^{(NMS-1)}\tilde{v}_{k,i}^{(NMS-2)}\prod_{\substack{q=1\\(q,j)\neq(i,M-1)}}^{N}\prod_{j=0}^{NMS-2}\sum_{h=1}^{M}\tilde{v}_{h,q}^{(j)}p_{hl}\right) =$$

$$\sum_{i=1}^{N}2\left(\tilde{v}_{k,i}^{(0)} + ... + \tilde{v}_{k,i}^{(NMS-2)}\right)\prod_{q=1}^{N}\prod_{j=0}^{NMS-2}\sum_{h=1}^{M}\tilde{v}_{h,q}^{(j)}p_{hl} \Rightarrow \quad \text{(S33)}$$

$$\sum_{i=1}^{N}\sum_{r=1}^{NMS-1}v_{l,i}^{(r)}\tilde{v}_{k,i}^{(r-1)}\prod_{\substack{q=1\\(q,j)\neq(i,r-1)}}^{N}\prod_{j=0}^{NMS-2}\sum_{h=1}^{M}\tilde{v}_{h,q}^{(j)}p_{hl} = 2\sum_{i=1}^{N}\sum_{r=1}^{NMS-1}\tilde{v}_{k,i}^{(r-1)}\prod_{q=1}^{N}\prod_{j=0}^{NMS-2}\sum_{h=1}^{M}\tilde{v}_{h,q}^{(j)}p_{hl}$$

Eq. (S33) in matrix form can be written as

$$\sum_{i=1}^{N} \left( \begin{bmatrix} \tilde{v}_{k,i}^{(0)} & \tilde{v}_{k,i}^{(1)} & \cdots & \tilde{v}_{k,i}^{(NMS-2)} \end{bmatrix} \begin{bmatrix} v_{l,i}^{(1)} \prod_{\substack{q=1 \\ (q,j) \neq (i,0)}}^{N} \prod_{j=0}^{NMS-2} \sum_{h=1}^{M} \tilde{v}_{h,q}^{(j)} p_{hl} \\ v_{l,i}^{(2)} \prod_{\substack{q=1 \\ (q,j) \neq (i,1)}}^{N} \prod_{j=0}^{NMS-2} \sum_{h=1}^{M} \tilde{v}_{h,q}^{(j)} p_{hl} \\ \vdots \\ v_{l,i}^{(NMS-1)} \prod_{\substack{q=1 \\ (q,j) \neq (i,NMS-2)}}^{N} \prod_{j=0}^{NMS-2} \sum_{h=1}^{M} \tilde{v}_{h,q}^{(j)} p_{hl} \end{bmatrix} \right) =$$

$$\sum_{i=1}^{N} \left( \begin{bmatrix} \tilde{v}_{k,i}^{(0)} & \tilde{v}_{k,i}^{(1)} & \cdots & \tilde{v}_{k,i}^{(NMS-2)} \end{bmatrix} \begin{bmatrix} 2 \prod_{q=1}^{N} \prod_{j=0}^{NMS-2} \sum_{h=1}^{M} \tilde{v}_{h,q}^{(j)} p_{hl} \\ 2 \prod_{q=1}^{N} \prod_{j=0}^{NMS-2} \sum_{h=1}^{M} \tilde{v}_{h,q}^{(j)} p_{hl} \\ \vdots \\ 2 \prod_{q=1}^{N} \prod_{j=0}^{NMS-2} \sum_{h=1}^{M} \tilde{v}_{h,q}^{(j)} p_{hl} \end{bmatrix} \right) \quad \text{(S34)}$$

Following the same direction as in Section C, we enforce the above constraint for each sample. We have

$$\begin{bmatrix} \tilde{v}_{k,i}^{(0)} & \tilde{v}_{k,i}^{(1)} & \cdots & \tilde{v}_{k,i}^{(NMS-2)} \end{bmatrix} \begin{bmatrix} v_{l,i}^{(1)} \prod_{\substack{q=1 \\ (q,j) \neq (i,0)}}^{N} \prod_{j=0}^{NMS-2} \sum_{h=1}^{M} \tilde{v}_{h,q}^{(j)} p_{hl} \\ v_{l,i}^{(2)} \prod_{\substack{q=1 \\ (q,j) \neq (i,1)}}^{N} \prod_{j=0}^{NMS-2} \sum_{h=1}^{M} \tilde{v}_{h,q}^{(j)} p_{hl} \\ \vdots \\ v_{l,i}^{(NMS-1)} \prod_{\substack{q=1 \\ (q,j) \neq (i,NMS-2)}}^{N} \prod_{j=0}^{NMS-2} \sum_{h=1}^{M} \tilde{v}_{h,q}^{(j)} p_{hl} \end{bmatrix} = \begin{bmatrix} \tilde{v}_{k,i}^{(0)} & \tilde{v}_{k,i}^{(1)} & \cdots & \tilde{v}_{k,i}^{(NMS-2)} \end{bmatrix} \begin{bmatrix} 2 \prod_{q=1}^{N} \prod_{j=0}^{NMS-2} \sum_{h=1}^{M} \tilde{v}_{h,q}^{(j)} p_{hl} \\ 2 \prod_{q=1}^{N} \prod_{j=0}^{NMS-2} \sum_{h=1}^{M} \tilde{v}_{h,q}^{(j)} p_{hl} \\ \vdots \\ 2 \prod_{q=1}^{N} \prod_{j=0}^{NMS-2} \sum_{h=1}^{M} \tilde{v}_{h,q}^{(j)} p_{hl} \end{bmatrix} \quad \text{(S35)}$$

Hence, for all values of $k$, we have the following equations,

$$\begin{bmatrix} \tilde{v}_{1,i}^{(0)} & \tilde{v}_{1,i}^{(1)} & \cdots & \tilde{v}_{1,i}^{(NMS-2)} \\ \tilde{v}_{2,i}^{(0)} & \tilde{v}_{2,i}^{(1)} & \cdots & \tilde{v}_{2,i}^{(NMS-2)} \\ \vdots & \vdots & \ddots & \vdots \\ \tilde{v}_{M,i}^{(0)} & \tilde{v}_{M,i}^{(1)} & \cdots & \tilde{v}_{M,i}^{(NMS-2)} \end{bmatrix} \left( \begin{bmatrix} v_{l,i}^{(1)} \prod_{\substack{q=1 \\ (q,j) \neq (i,0)}}^{N} \prod_{j=0}^{NMS-2} \sum_{h=1}^{M} \tilde{v}_{h,q}^{(j)} p_{hl} \\ v_{l,i}^{(2)} \prod_{\substack{q=1 \\ (q,j) \neq (i,1)}}^{N} \prod_{j=0}^{NMS-2} \sum_{h=1}^{M} \tilde{v}_{h,q}^{(j)} p_{hl} \\ \vdots \\ v_{l,i}^{(NMS-1)} \prod_{\substack{q=1 \\ (q,j) \neq (i,NMS-2)}}^{N} \prod_{j=0}^{NMS-2} \sum_{h=1}^{M} \tilde{v}_{h,q}^{(j)} p_{hl} \end{bmatrix} - \begin{bmatrix} 2 \prod_{q=1}^{N} \prod_{j=0}^{NMS-2} \sum_{h=1}^{M} \tilde{v}_{h,q}^{(j)} p_{hl} \\ 2 \prod_{q=1}^{N} \prod_{j=0}^{NMS-2} \sum_{h=1}^{M} \tilde{v}_{h,q}^{(j)} p_{hl} \\ \vdots \\ 2 \prod_{q=1}^{N} \prod_{j=0}^{NMS-2} \sum_{h=1}^{M} \tilde{v}_{h,q}^{(j)} p_{hl} \end{bmatrix} \right) = \mathbf{0}_{M \times 1}. \quad \text{(S36)}$$

If the matrix $\tilde{\mathbf{V}}_i$ given by

$$\tilde{\mathbf{V}}_i = \begin{bmatrix} \tilde{v}_{1,i}^{(0)} & \tilde{v}_{1,i}^{(1)} & \cdots & \tilde{v}_{1,i}^{(NMS-2)} \\ \tilde{v}_{2,i}^{(0)} & \tilde{v}_{2,i}^{(1)} & \cdots & \tilde{v}_{2,i}^{(NMS-2)} \\ \vdots & \vdots & \ddots & \vdots \\ \tilde{v}_{M,i}^{(0)} & \tilde{v}_{M,i}^{(1)} & \cdots & \tilde{v}_{M,i}^{(NMS-2)} \end{bmatrix} \quad (S37)$$

is full column rank, its null space is reduced to the singleton $\{0\}$. This matrix is column full rank when the number of column is less than the number of rows, i.e., $NMS - 1 < M$, which means $\mathbf{V}_i^T \mathbf{V}_i$ is invertible. Hence, by this assumption we have

$$\begin{bmatrix} v_{l,i}^{(1)} \prod_{\substack{q=1 \\ (q,j) \neq (i,0)}}^{N} \prod_{j=0}^{NMS-2} \sum_{h=1}^{M} v_{h,q}^{(j)} p_{hl} \\ v_{l,i}^{(2)} \prod_{\substack{q=1 \\ (q,j) \neq (i,1)}}^{N} \prod_{j=0}^{NMS-2} \sum_{h=1}^{M} v_{h,q}^{(j)} p_{hl} \\ \vdots \\ v_{l,i}^{(NMS-1)} \prod_{\substack{q=1 \\ (q,j) \neq (i,NMS-2)}}^{N} \prod_{j=0}^{NMS-2} \sum_{h=1}^{M} v_{h,q}^{(j)} p_{hl} \end{bmatrix} - \begin{bmatrix} 2 \prod_{q=1}^{N} \prod_{j=0}^{NMS-2} \sum_{h=1}^{M} \tilde{v}_{h,q}^{(j)} p_{hl} \\ 2 \prod_{q=1}^{N} \prod_{j=0}^{NMS-2} \sum_{h=1}^{M} \tilde{v}_{h,q}^{(j)} p_{hl} \\ \vdots \\ 2 \prod_{q=1}^{N} \prod_{j=0}^{NMS-2} \sum_{h=1}^{M} \tilde{v}_{h,q}^{(j)} p_{hl} \end{bmatrix} = \mathbf{0}_{M \times 1}. \quad (S38)$$

Simplifying (S38), we have

$$\begin{bmatrix} v_{l,i}^{(1)} \\ v_{l,i}^{(2)} \\ \vdots \\ v_{l,i}^{(NMS-1)} \end{bmatrix} = \begin{bmatrix} 2 \sum_{h=1}^{M} \tilde{v}_{h,i}^{(0)} p_{hl} \\ 2 \sum_{h=1}^{M} \tilde{v}_{h,i}^{(1)} p_{hl} \\ \vdots \\ 2 \sum_{h=1}^{M} \tilde{v}_{h,i}^{(NMS-2)} p_{hl} \end{bmatrix} \Rightarrow$$

$$\begin{bmatrix} v_{l,i}^{(1)} \\ v_{l,i}^{(2)} \\ \vdots \\ v_{l,i}^{(NMS-1)} \end{bmatrix} = 2 \begin{bmatrix} \tilde{v}_{1,i}^{(0)} & \tilde{v}_{2,i}^{(0)} & \cdots & \tilde{v}_{M,i}^{(0)} \\ \tilde{v}_{1,i}^{(1)} & \tilde{v}_{2,i}^{(1)} & \cdots & \tilde{v}_{M,i}^{(1)} \\ \vdots & \vdots & \ddots & \vdots \\ \tilde{v}_{1,i}^{(NMS-2)} & \tilde{v}_{2,i}^{(NMS-2)} & \cdots & \tilde{v}_{M,i}^{(NMS-2)} \end{bmatrix} \begin{bmatrix} p_{1l} \\ p_{2l} \\ \vdots \\ p_{Ml} \end{bmatrix} \quad (S39)$$

This is the model we use for the Poisson observations of the cell proliferation Markov process.

**Section E. Proof of Proposition 1**

Starting from (S39), since a direction solution for the transition probabilities is not accessible in this case, we resort to an approximation. To this end, we use an one sample estimator for $\tilde{v}_{1,i}^{(0)}$ and replace it by $v_{1,i}^{(0)}$. We have

$$2\begin{bmatrix} v_{1,i}^{(0)} & v_{2,i}^{(0)} & \cdots & v_{M,i}^{(0)} \\ v_{1,i}^{(1)} & v_{2,i}^{(1)} & \cdots & v_{M,i}^{(1)} \\ \vdots & \vdots & \ddots & \vdots \\ v_{1,i}^{(NMS-2)} & v_{2,i}^{(NMS-2)} & \cdots & v_{M,i}^{(NMS-2)} \end{bmatrix} \begin{bmatrix} p_{1l} \\ p_{2l} \\ \vdots \\ p_{Ml} \end{bmatrix} = \begin{bmatrix} v_{l,i}^{(1)} \\ v_{l,i}^{(2)} \\ \vdots \\ v_{l,i}^{(NMS-1)} \end{bmatrix}. \quad \text{(S40)}$$

where, $i \in \{1, N\}$. The equation (S40) is satisfied for each sample, $i$, and cell type, $l$. The goal is set to find $p_{hl}$, $h, l \in [1, M]$, which fit the equations best in the minimum mean squared sense. We have

$$\hat{\mathbf{P}} = \arg\min_{\mathbf{P}} S(\mathbf{P}) \quad \text{(S41-a)}$$

s.t.

for $\forall \ 1 \leq h \leq M, 1 \leq l \leq M$,

$$0 \leq p_{hl} \leq 1, \quad \text{(S41-b)}$$

$$\sum_{l=1}^{M} p_{hl} = 1, \quad \text{(S41-c)}$$

where the objective function $S$ is defined as

$$S(\mathbf{P}) = \sum_{l=1}^{M} \left( \sum_{i=1}^{N} \sum_{j=1}^{NMS-1} \left( v_{l,i}^{(j)} - 2 \sum_{h=1}^{M} \tilde{v}_{h,i}^{(j-1)} p_{hl} \right)^2 \right), \quad \text{(S42)}$$

The optimization problem in (S41) is convex, (cost function is quadratic and constraints are linear [2]). Using KKT (Karush–Kuhn–Tucker), the unconstrained solution for this problem is given by

$$S_U(\mathbf{P}) = \sum_{l=1}^{M} \left( S(\mathbf{P}_l) + \lambda_l \left( \sum_{j=1}^{M} p_{jl} - 1 \right) + \sum_{j=1}^{M} (\lambda'_{jl} - \lambda''_{jl}) p_{jl} \right), \quad \text{(S43)}$$

where $\lambda'_{jl}$, $\lambda''_{jl}$ and $\lambda_l$ are Lagrange multipliers and $\mathbf{P}_l = [p_{1l} \ \cdots \ p_{Ml}]$. Computing the derivative of the cost function, with respect to $p_{hl}$ and following some mathematical manipulations (similar to the Gaussian scenario), we have

$$\hat{\mathbf{P}}_l = 0.5 \left( \left( \mathbf{V}_{:,1:N}^{(0:NMS-2)} \right)^\dagger \left( \mathbf{V}_{:,1:N}^{(0:NMS-2)} \right) \right)^{-1} \left( \mathbf{V}_{:,1:N}^{(0:NMS-2)} \right)^\dagger \mathbf{V}_{l,(1:N)}^{(1:NMS-1)} - \Lambda_l, \quad \text{(S44)}$$

where $\mathbf{V}_{:,1:N}^{(0:NMS-2)}$, $\mathbf{V}_{l,(1:N)}^{(1:NMS-1)}$ and $\Lambda_l$ are defined in (S-13), (S-14) and (S-16). The above solution is obtained when the constraints $(0 \leq p_{ij} \leq 1)$ is assumed satisfied. However, if the obtained results violate these constraints, one should enforce them and solve the problem again (see Section B for details on handling this issue).

## Section F. Proof of Proposition 2

Starting from (S39), since a direction solution for the transition probabilities is not accessible in this case, we resort to an alternative approximation in this Section. To this end, we consider the summation of (S39) over $i$ and obtain

$$2\begin{bmatrix} \sum_{i=1}^{N} \tilde{v}_{1,i}^{(0)} & \cdots & \sum_{i=1}^{N} \tilde{v}_{M,i}^{(0)} \\ \vdots & \ddots & \vdots \\ \sum_{i=1}^{N} \tilde{v}_{1,i}^{(NMS-2)} & \cdots & \sum_{i=1}^{N} \tilde{v}_{M,i}^{(NMS-2)} \end{bmatrix} \begin{bmatrix} p_{1l} \\ \vdots \\ p_{Ml} \end{bmatrix} = \begin{bmatrix} \sum_{i=1}^{N} v_{l,i}^{(1)} \\ \vdots \\ \sum_{i=1}^{N} v_{l,i}^{(NMS-1)} \end{bmatrix}. \quad (S45)$$

Using the definition of a sample mean, and assuming $\frac{1}{N}\sum_{i=1}^{N} \tilde{v}_{l,i}^{k} = \frac{1}{N}\sum_{i=1}^{N} v_{l,i}^{k}$, we have

$$2\begin{bmatrix} \sum_{i=1}^{N} v_{1,i}^{(0)} & \cdots & \sum_{i=1}^{N} v_{M,i}^{(0)} \\ \vdots & \ddots & \vdots \\ \sum_{i=1}^{N} v_{1,i}^{(NMS-2)} & \cdots & \sum_{i=1}^{N} v_{M,i}^{(NMS-2)} \end{bmatrix} \begin{bmatrix} p_{1l} \\ \vdots \\ p_{Ml} \end{bmatrix} = \begin{bmatrix} \sum_{i=1}^{N} v_{l,i}^{(1)} \\ \vdots \\ \sum_{i=1}^{N} v_{l,i}^{(NMS-1)} \end{bmatrix}. \quad (S46)$$

If $NMS - 1 < M$, this linear algebraic system has an infinite number of solutions. If $NMS - 1 = M$, the above system of linear equations for $p_{1l}$ has a single solution

$$\begin{bmatrix} p_{1l} \\ \vdots \\ p_{Ml} \end{bmatrix} = 0.5 \begin{bmatrix} \sum_{i=1}^{N} v_{1,i}^{(0)} & \cdots & \sum_{i=1}^{N} v_{M,i}^{(0)} \\ \vdots & \ddots & \vdots \\ \sum_{i=1}^{N} v_{1,i}^{(NMS-2)} & \cdots & \sum_{i=1}^{N} v_{M,i}^{(NMS-2)} \end{bmatrix}^{-1} \begin{bmatrix} \sum_{i=1}^{N} v_{l,i}^{(1)} \\ \vdots \\ \sum_{i=1}^{N} v_{l,i}^{(NMS-1)} \end{bmatrix} \quad (S47)$$

If $NMS - 1 > M$, the solution in MSE sense is obtained with the same approach for the Gaussian distribution by replacing $\mathbf{V}_{:,1:N}^{(0:NMS-2)}$ and $\mathbf{V}_{l,(1:N)}^{(1:NMS-1)}$ in (S13) and (S14) by $\mathbf{\Sigma}_v$ and $\Psi_l$, which are described as follows

$$\mathbf{\Sigma}_v = \begin{bmatrix} \sum_{i=1}^{N} v_{1,i}^{(0)} & \cdots & \sum_{i=1}^{N} v_{M,i}^{(0)} \\ \vdots & \ddots & \vdots \\ \sum_{i=1}^{N} v_{1,i}^{(NMS-2)} & \cdots & \sum_{i=1}^{N} v_{M,i}^{(NMS-2)} \end{bmatrix}. \quad (S48)$$

$$\Psi_l = \begin{bmatrix} \sum_{i=1}^{N} v_{l,i}^{(1)} & \cdots & \sum_{i=1}^{N} v_{l,i}^{(NMS-1)} \end{bmatrix}. \quad (S49)$$

The above solution is obtained when the constraints $\left(0 \leq p_{ij} \leq 1\right)$ is assumed satisfied. However, if the obtained results violate these constraints, one should enforce them and solve the problem again (see Section B for details on handling this issue).

References

1	Maletzki, C., Stier, S., Gruenert, U., Gock, M., Ostwald, C., Prall, F., and Linnebacher, M.: 'Establishment, Characterization and Chemosensitivity of Three Mismatch Repair Deficient Cell Lines from Sporadic and Inherited Colorectal Carcinomas', Plos One, 2012, 7, (12)
2	Boyd, S.P., and Vandenberghe, L.: 'Convex optimization' (Cambridge University Press, 2004. 2004)
3	Nocedal, J., and Wright, S.J.: 'Numerical optimization' (Springer, 2006, 2nd edn. 2006) pp. 314-357